\documentclass[journal, letterpaper,10pt,twocolumn]{IEEEtran}

\usepackage[T1]{fontenc}
\usepackage[english]{babel}

\usepackage{csquotes} %
\usepackage{tikz}
\usepackage{tikzscale}
\usepackage{pgfplots}
\pgfplotsset{compat=1.17}
\usepackage{tabularx}
\usepackage{booktabs}
\usepackage{multirow}
\usepackage{pgfplotstable}
\usepackage[toc,nopostdot]{glossaries}
\usepackage{ifthen}
\usepackage{enumitem}

\usetikzlibrary{external}
\usepackage{mathtools}
\usepackage{amsthm}
\usepackage{amssymb} %
\usepackage{siunitx}
\usepackage{bm}
\usepackage[ruled,english,onelanguage,lined]{algorithm2e}
\usepackage[super]{nth}
\usepackage{trfsigns} %

\usepackage{xspace}
\usepackage{pdfpages}
\usepackage{graphicx}

\usetikzlibrary{pgfplots.groupplots} %
\usetikzlibrary{backgrounds}
\usetikzlibrary{arrows.meta}
\usetikzlibrary{calc}
\usetikzlibrary{spy}
\usetikzlibrary{plotmarks}

\newcommand*{\fig}[1]{Fig.~\ref{#1}}
\newcommand*{\tab}[1]{Tab.~\ref{#1}}

\makeatletter
\DeclareRobustCommand\onedot{\futurelet\@let@token\@onedot}
\def\@onedot{\ifx\@let@token.\else.\null\fi\xspace}

\def\eg{e.g\onedot} 
\def\ie{i.e\onedot} 
\def\cf{c.f\onedot} 
 
\def\wrt{w.r.t\onedot} 
\def\etal{et al\onedot}
\makeatother

\newcommand*{\mr}[1]{\mathrm{#1}}
\newcommand*{\complexityupper}[1]{\mathcal{O}\left(#1\right)}
\newcommand*{\matset}[1]{\left\{#1\right\}}

\newcommand*{\inv}{^{-1}}

\newcommand*{\abs}[1]{\left\lvert{#1}\right\rvert}

\newcommand*{\real}[1]{\Re\left\{#1\right\}}

\newcommand*{\itindex}[1]{^{(#1)}} %

\newcommand*{\vect}[1]{\boldsymbol{\mathbf{#1}}} %
\newcommand*{\matr}[1]{\boldsymbol{\mathbf{#1}}} %
\newcommand*{\tran}{^{\operatorname{T}}}
\newcommand*{\herm}{^{\operatorname{H}}}
\newcommand*{\nullmat}{\matr{0}}
\newcommand*{\onevec}{\vect{1}}
\newcommand*{\nullvec}{\vect{0}}
\newcommand*{\identity}{\matr{I}}

\newcommand*{\fronorm}[1]{{\left\lVert#1\right\rVert}_F}
\newcommand*{\trace}[1]{\operatorname{Tr}\left\{#1\right\}}

\newcommand*{\vecbrac}[1]{\left[#1\right]} %
\newcommand*{\logdet}[1]{\operatorname{logdet}\left(#1\right)}
\newcommand*{\matel}[2]{\left[#1\right]_{#2}}

\newcommand*{\prob}[1]{\mathop{}\!p(#1)}
\newcommand*{\expectation}[2]{\mathbb{E}_{#1}\left[#2\right]}

\newcommand*{\normdistr}[1]{\mathcal{N}\left(#1\right)}
\newcommand*{\cnormdistr}[1]{\mathcal{C}\normdistr{#1}}

\DeclareMathOperator*{\argmax}{argmax}
\newcommand*{\subjectto}{\text{subject to}}

\newcommand*\Rset{\mathbb{R}}
\newcommand*\Cset{\mathbb{C}}

\DeclareMathOperator*{\relu}{ReLU}

\DeclareMathOperator*{\modrelu}{{modReLU}}

\DeclareSIUnit[per-mode=symbol,per-symbol=p]{\MHz}{\mega \hertz}
\DeclareSIUnit[per-mode=symbol,per-symbol=p]{\GHz}{\giga \hertz}
\DeclareSIUnit{\belmilliwatt}{Bm}
\DeclareSIUnit{\dBm}{\deci\belmilliwatt}
\DeclareSIUnit[per-mode=symbol,per-symbol=p]{\perhertz}{\hertz^{-1}}
\DeclareSIUnit[per-mode=symbol,per-symbol=p]{\permeter}{\meter^{-1}}
\DeclareSIUnit[per-mode=symbol,per-symbol=p]{\meterpersecond}{\meter \second^{-1}}
\DeclareSIUnit[]{\nat}{nat}
\DeclareSIUnit[per-mode=fraction]{\nats}{\nat \per \hertz}

\definecolor{red}{HTML}{cc2900}
\definecolor{blue}{HTML}{0E84B2}
\definecolor{green}{HTML}{3cb44b}
\definecolor{orange}{HTML}{f58231}
\definecolor{purple}{HTML}{984ea3}
\definecolor{cyan}{HTML}{00FFFF}
\definecolor{fred}{HTML}{FF0000} %

\definecolor{green1}{HTML}{bae4b3}
\definecolor{green2}{HTML}{74c476}
\definecolor{green3}{HTML}{31a354}
\definecolor{green4}{HTML}{006d2c}
\definecolor{blue1}{HTML}{bae4b3}
\definecolor{blue2}{HTML}{6baed6}
\definecolor{blue3}{HTML}{3182bd}
\definecolor{blue4}{HTML}{08519c}
\definecolor{red1}{HTML}{ffad99}
\definecolor{red2}{HTML}{ff5c33}
\definecolor{red3}{HTML}{cc2900}
\definecolor{red4}{HTML}{661400}
\definecolor{red5}{HTML}{000000}

\pgfplotscreateplotcyclelist{markcycle}{mark = o, mark = triangle, mark = diamond, mark = square, mark = pentagon, mark = star, mark = oplus, mark = otimes}

\pgfplotsset{
	minor grid style={dotted},
	major grid style={dotted},
	grid = major, 
	cycle list name = markcycle,
	every axis plot/.append style={thick, mark options=solid,},
	legend style={at={(1,1.03)}, draw=none, fill=none, font=\small, anchor=south east, /tikz/every even column/.append style={column sep=0.2cm}},
}

\usepackage[style=ieee, backend=biber]{biblatex}

\usetikzlibrary{external}
\newcommand*{\rate}{\mathcal{R}}
\newcommand*{\vvtilde}{\tilde{\vect{v}}}
\newcommand*{\vmtilde}{\tilde{\matr{V}}}
\newcommand*{\vm}{{\matr{V}}}
\newcommand*{\vv}{{\vect{v}}}

\newcommand*{\um}{{\matr{U}}}

\newcommand*{\wm}{{\matr{W}}}

\newcommand*{\rmat}{{\matr{R}}}

\newcommand*{\hmat}{{\matr{H}}}

\newcommand*{\setik}{{\mathcal{I}_k}}

\newacronym{bs}{BS}{base station}
\newacronym{ue}{UE}{user equipment}
\newacronym{snr}{SNR}{signal-to-noise ratio}
\newacronym{sinr}{SINR}{signal-to-noise-plus-interference ratio}
\newacronym{slnr}{SLNR}{signal-to-leakage-plus-noise ratio}
\newacronym{bch}{BCH}{broadcast channel}
\newacronym{ifc}{IFC}{interference channel}
\newacronym{mu}{MU}{multi-user}
\newacronym{siso}{SISO}{single-input single-output}
\newacronym{miso}{MISO}{multiple-input single-output}
\newacronym{mimo}{MIMO}{multiple-input multiple-output}
\newacronym{csi}{CSI}{channel state information}
\newacronym{wsr}{WSR}{weighted sum-rate}
\newacronym{mrc}{MRC}{maximum-ratio combining}
\newacronym{zf}{ZF}{zero-forcing}
\newacronym{rzf}{RZF}{regularized zero-forcing}
\newacronym{mse}{MSE}{mean squared error}
\newacronym{mmse}{MMSE}{minimum mean squared error}
\newacronym{bd}{BD}{block diagonalization}
\newacronym{wmmse}{WMMSE}{weighted minimum mean squared error}
\newacronym{kkt}{KKT}{Karush-Kuhn-Tucker}
\newacronym{bcd}{BCD}{block coordinate descent}
\newacronym{sgd}{SGD}{stochastic gradient descent}
\newacronym{mlp}{MLP}{multi-layer perceptron}
\newacronym{gcf}{GCF}{graph convolutional filter}
\newacronym{gnn}{GNN}{graph neural network}
\newacronym{gcn}{GCN}{graph convolutional network}
\newacronym{arma}{ARMA}{autoregressive moving average}
\newacronym{cnn}{CNN}{convolutional neural network}
\newacronym{rnn}{RNN}{reocurrent neural network}
\newacronym{psd}{PSD}{positive semidefinite}
\newacronym{cg}{CG}{conditional gradient}
\newacronym{pgd}{PGD}{projected gradient descent}
\newacronym{qp}{QP}{quadratic program}
\newacronym{cpu}{CPU}{central processing unit}
\newacronym{gpu}{GPU}{graphics processing unit}
\newacronym{nan}{NaN}{Not-a-Number}
\newacronym{slurm}{Slurm}{Slurm Workload Manager}
\newacronym{qcqp}{QCQP}{quadratically constrained quadratic program}
\newacronym{dl}{DL}{downlink}

\usepackage{hyperref}
\newcommand\copyrighttext{%
	\footnotesize \textcopyright 2023 IEEE. Personal use of this material is permitted.
	Permission from IEEE must be obtained for all other uses, in any current or future
	media, including reprinting/republishing this material for advertising or promotional
	purposes, creating new collective works, for resale or redistribution to servers or
	lists, or reuse of any copyrighted component of this work in other works.
	DOI: \href{https://ieeexplore.ieee.org/document/10040241}{10.1109/JSAC.2023.3242716}}

\makeatletter
\def\ps@IEEEtitlepagestyle{
	\def\@oddfoot{\mycopyrightnotice}
	\def\@evenfoot{}
}
\def\mycopyrightnotice{
	{\footnotesize
		\begin{minipage}{\textwidth-2\fboxsep}%
		\centering%
		\noindent\fbox{\parbox{\linewidth}{\copyrighttext}}
		\end{minipage}
	}
}

\begin{document}
	\title{Coordinated Sum-Rate Maximization in Multicell MU-MIMO with Deep Unrolling}
	\author{\IEEEauthorblockN{Lukas~Schynol},~\and \IEEEauthorblockN{Marius Pesavento}%
	\thanks{The authors acknowledge the financial support by the Federal Ministry of Education and Research of Germany in the project "Open6GHub" (grant no. 16KISK014). The computations for this research paper were conducted on the Lichtenberg high performance computer of the TU Darmstadt. This article was presented in part at the 30th European Signal Processing Conference (EUSIPCO), 2022.}%
	\thanks{Lukas Schynol (corresponding author) and Marius Pesavento are with the Communication Systems Group, Technische Universit\"at Darmstadt, Darmstadt, Germany; Emails: \{lschynol, pesavento\}@nt.tu-darmstadt.de}%
	\thanks{Accepted for publication in IEEE Journal on Selected Areas in Communications.}%
	}%

	\maketitle
	\begin{abstract}
		Coordinated weighted sum-rate maximization in multicell MIMO networks with intra- and intercell interference and local channel state at the base stations is recognized as an important yet difficult problem.
		A classical, locally optimal solution is obtained by the \gls{wmmse} algorithm which facilitates a distributed implementation in multicell networks.
		However, it often suffers from slow convergence and therefore large communication overhead.
		To obtain more practical solutions, the unrolling/unfolding of traditional iterative algorithms gained significant attention.
		In this work, we demonstrate a complete unfolding of the \gls{wmmse} algorithm for transceiver design in multicell \acrshort{mu}-\acrshort{mimo} interference channels with local channel state information.
		The resulting architecture termed GCN-WMMSE applies ideas from graph signal processing and is agnostic to different wireless network topologies, while exhibiting a low number of trainable parameters and high efficiency \wrt training data.
		It significantly reduces the number of required iterations while achieving performance similar to the \gls{wmmse} algorithm, alleviating the overhead in a distributed deployment.
		Additionally, we review previous architectures based on unrolling the \gls{wmmse} algorithm and compare them to GCN-WMMSE in their specific applicable domains.
	\end{abstract}
	\glsresetall
	

	\section{Introduction}
	
	\par
	
	\Gls{mimo} system theory has been instrumental in meeting the spectral efficiency and capacity targets of modern wireless systems \cite{heath_lozano_2018_foundations_of_mimo}.
	When considering downlink beamforming with instantaneous \gls{csi}, a common utility function to design optimal transmit and receive beamformers given transmit power limitations is the constrained \gls{wsr} maximization.
	Treating interference as noise for simple receivers, the objective is multimodal and its optimization is NP-hard, even for the single-antenna case or a single-cell network \cite{lou2008_spectrum_management,hunger2008_combinatorial_approach_mimobc_nonconvex}.
	Globally optimal algorithms such as \cite{bjoernson2011_optimal_miso_multicell} were developed but consequently require exponential runtime.
	In the space of locally optimal algorithms, the \gls{wmmse} iterative algorithm \cite{shi2011_wmmse} is widely regarded as a benchmark for \gls{wsr} maximization due to the high sum-rate it attains while its updates can be expressed in closed-form.
	Furthermore, it is applicable in multicell systems and can be deployed in a distributed fashion with limited communication overhead per iteration, eliminating the necessity to collect the total channel state information at a central point.
	Depending on the channel conditions, the \gls{wmmse} algorithm requires a large number of iterations to converge.
	The authors of \cite{shen2018_fractional_programming} proposed an algorithm based on fractional programming which converges in fewer iterations than the \gls{wmmse} algorithm. However, it adds significant complexity by introducing an additional convex subproblem per iteration which itself must be iteratively solved.
	Nevertheless, due to their associated runtime or communication overhead, optimization-based iterative solutions are difficult to apply in practice.
	\par
	In recent years, machine learning methods have been identified as one of the key components for the design and operation of future communication networks \cite{peltonen_etal2020_6g_white_paper, bjoernson2019_massive_mimo_reality, zhang2020_6g_ai_survey}.
	Among other use cases such as resource management \cite{challita2017_resource_management_dl}, detection \cite{samuel2017_deep_mimo_detection} or channel estimation \cite{neumann2017_deep_channel_estimation}, the application of deep learning on transmit power allocation and downlink beamforming has been explored \cite{sun2018_dnn_interference_management, alkhateeb2018_deeplearned_coordinatedbf, eisen2020_regnn_wireless_resource_allocation, xia2020_miso_singlecell_beamforming_nn, huang_cnn_beamforming}.
	Particularly in beamforming, neural networks exhibit promising results w.r.t. their achieved sum-rate performance compared to their computational complexity.
	\par
	First explicitly explored in \cite{gregor2010_unfolding_lista}, the concept of algorithm unfolding or algorithm unrolling recently gained significant interest in the signal processing community \cite{monga2020_algorithm_unrolling, balatsoukasstimming2019_deepunrolling_comm}.
	While many state-of-the-art techniques incorporate components of classical algorithms into their corresponding deep network architecture, algorithm unrolling takes this notion further by viewing iterations of these problem-specific algorithms as layers of machine learning models.
	Given that the function representing the objective variable update in one iteration can be made fully differentiable, we can unfold a finite number of iterations, introduce new parameters and train them on data using established \gls{sgd} techniques.
	This way of combining expert knowledge with machine learning concepts can lead to better generalization performance and interpretability of unrolling-based architectures compared to conventional neural network architectures.
	These two factors are key for robust wireless communication networks, and together with the already existing body of algorithms based on rigorous optimization, algorithm unrolling is an attractive alternative to the use of generic network architectures.
	\par
	Several works applied the concept of algorithm unrolling to the problem of downlink beamforming.
	In \cite{zhu2020_learning_beamforming_unfolding_pgd}, efficient solutions to the \gls{wsr} maximization in \gls{miso} networks consisting of transmitter-receiver pairs are found by unrolling the inexact cyclic coordinate descent method \cite{liu2011_inexact_cyclic_coordinate_descent}.
	In \cite{pellaco2020_deep_unfolding_pgd}, a substep of the \gls{wmmse} algorithm is replaced by a secondary \gls{pgd} optimization, the step sizes of which are learned using data, thereby facilitating a trade-off between performance and complexity. 
	However, the discussion in \cite{pellaco2020_deep_unfolding_pgd} is limited to single-cell \gls{miso} networks.
	The architecture proposed in \cite{hu2021_iadnn} approximates and unfolds matrix inversion operations within the updates of the \gls{wmmse} algorithm, achieving performance close to the \gls{wmmse} in the single-cell \gls{mimo} case with multiple users (MU).
	The authors of \cite{chowdhury2021_unfolding} consider a multicell scenario of pairwise \gls{siso} links and incorporate \glspl{gnn} \cite{zhou2018_gnn_review_method_apps} into the \gls{wmmse} variable updates in order to reduce the number of required iterations. 
	\par
	In this paper, the structure of \glspl{gcn} is applied to unfold the \gls{wmmse} algorithm in its most general form, i.e., coordinated multicell \gls{mu}-\gls{mimo} networks.
	Unlike other approaches \cite{eisen2020_regnn_wireless_resource_allocation, chowdhury2021_unfolding, shen2021_gnn_wireless_resource} based on \glspl{gnn}, in which the graph underlying the \gls{gnn} mimicks the wireless network topology, our \gls{gnn} graph shift matrices represent transceiver signal spaces to achieve generality in the number of transceiver antennas.
	To our best knowledge, transceiver design algorithms based on deep unrolling have not yet been considered in general network scenarios including \gls{mu}-\gls{mimo} cells with inter- and intracell interference.
	We set the following requirements and goals for our deep unfolded network:
	\begin{itemize}
		\item Compared to previous literature, we focus on a machine learning architecture that is at least as general as the original WMMSE algorithm, and that is robust to changing network configurations after training.
		\item The number of iterations must reduce while the computational effort per iteration must not exceed the complexity order of the original \gls{wmmse} algorithm.
		\item The communication overhead per iteration for distributed implementations must be equal or less compared to the WMMSE algorithm to capitalize on the reduced number of iterations.
	\end{itemize}
	Our contributions are:
	\begin{enumerate}
		\item Utilizing \glspl{gcn} as spatial signal filters, we propose a completely general unfolding architecture of the WMMSE algorithm  for \gls{wsr} maximization in multicell \gls{mu}-\gls{mimo} networks that meets the above requirements, which we denote as GCN-WMMSE.
		\item We propose an accelerated dual variable update procedure based on rational function approximation for the classical \gls{wmmse} algorithm and the forward pass of our GCN-WMMSE networks. %
		In the backward pass, we provide an efficient network parameter update scheme based on the notion of implicit functions.
		Furthermore, we pinpoint numerical issues that can arise for both methods.
		\item We demonstrate the excellent generalization capability and training data efficiency of the proposed GCN-WMMSE architecture in extensive simulations on Rayleigh fading and ray-tracing channel models. 
		The resulting flexibility is exploited to reduce the cost of communication and computation of model training.
		\item We draw extensive comparisons to previously published unfolding methods, and extend their results by further experiments, leading to insights useful for future designs.
	\end{enumerate}
	\par
	The rest of the paper is structured as follows: 
	Section \ref{sec:scenario_wmmse} defines the wireless system model and formulates the resource allocation problem, followed by a review of the \gls{wmmse} algorithm.
	The GCN-WMMSE architecture is proposed and motivated in Section \ref{sec:model}.
	Section \ref{sec:experiments} presents the simulations results.
	In Section \ref{sec:relatedwork}, we relate our proposed architecture to previous works on \gls{wmmse} algorithm unrolling and provide additional simulation results with numerical comparisons. 
	Section \ref{sec:conclusion} provides concluding remarks.
	\par
	\textit{Notation:} Uppercase $\matr{X}$ and lowercase $\vect{x}$ bold letters denote matrices and vectors respectively, where matrix elements are accessed by $\matel{\matr{X}}{ij}$. 
	$(.)\herm$,$(.)\tran$, $(.)\inv$ and $(.)^\dagger$ denote the Hermitian transpose, transpose, inverse and Moore-Penrose inverse respectively. 
	We denote the trace by $\trace{.}$, the Frobenius norm by $\fronorm{.}$ and the log-determinant \wrt the natural basis as $\logdet{.}$.
	$\identity$ is the identity matrix, $\onevec$ denotes a vector filled with ones.
	$\nabla_{\vect{x}}$ denotes the gradient w.r.t. to some $\vect{x}$. 
	$f'$ denotes the derivative of a function $f$ respectively.
	
	\section{Weighted MMSE Minimization} \label{sec:scenario_wmmse}
	\subsection{System Model and Downlink Problem Formulation}
	Consider a wireless cellular system consisting of $K$ cells, each of which are defined by a \gls{bs}.
	\gls{bs} $k$ is serving one of $K$ disjoint subsets $\matset{\mathcal{I}_k}_{k=1}^K$ of users or \glspl{ue}, for a total of $I = \sum_{k=1}^K \abs{\mathcal{I}_k}$ \glspl{ue}, and is equipped with an antenna array of size $M_k$, whereas every \gls{ue} $i$ is equipped with an array of $N_i$ antennas. 
	The output signal of the antenna array of \gls{bs} $k$ is modeled as the sum of mapped symbols $\vect{x}_{\mr{Tx},k} = \sum\nolimits_{i \in \mathcal{I}_k} \matr{V}_i \breve{\vect{s}}_{i}$ for $k =1, \dots, K$
	where $\matr{V}_i \in \Cset^{M_k \times N_i}$ is the complex downlink precoding matrix and $\breve{\vect{s}}_i \in \Cset ^{N_i}$ with $\expectation{}{\breve{\vect{s}}_i \breve{\vect{s}}_i\herm} = \identity$ is the symbol vector directed at \gls{ue} $i$ respectively\footnotemark.
	\footnotetext{Without loss of generality the number of data symbols per time instant per \gls{ue} $i$ is assumed to be equal to the number of receive antennas $N_i$.}
	We assume that the transceivers employ Gaussian code books and that the symbols are statistically independent between \glspl{ue}.
	Given complex frequency-flat channel matrices $\matr{H}_{ik} \in \Cset^{N_i\times M_k}$ from \gls{bs} $k$ to \gls{ue} $i$ and block-fading, the received signal of a user $i$ assigned to \gls{bs} $k$ becomes
	\begin{equation*}
		\setlength{\abovedisplayskip}{4pt}
		\setlength{\belowdisplayskip}{4pt}
		\vect{y}_{\mr{Rx},i} = 
		\matr{H}_{i k} \matr{V}_i \breve{\vect{s}}_i 
		+ \sum\nolimits_{\substack{m=1}}^K \sum\nolimits_{j \in \mathcal{I}_m \setminus \{i\} } \matr{H}_{i m} \matr{V}_j \breve{\vect{s}}_j + \vect{n}_i,
	\end{equation*}
	where $\vect{n}_i \sim \cnormdistr{0,\sigma_i^2 \identity}$ is additive complex white Gaussian noise with power $\sigma_i^2$ per antenna element. 
	The second term both contains the intracell and intercell interference.
	We treat interference as noise at the \glspl{ue}, thus, the achievable rate at \gls{ue} $i$ can be determined as $\rate_i = B \logdet{\identity + \matr{Q}_i \matr{Z}_i \inv}$ with $\matr{Q}_i = \matr{H}_{i k} \matr{V}_i \matr{V}_i\herm \matr{H}_{i k}\herm$ being the covariance matrix of the useful signal and $\matr{Z}_i = \sum_{m=1}^K \sum_{j \in \mathcal{I}_m \setminus \{i\} } \matr{H}_{i m} \matr{V}_j \matr{V}_j\herm \matr{H}_{i m} \herm + \sigma_i^2 \identity$ being the covariance matrix of the interference plus noise.
	Without loss of generality the bandwidth $B$ is set to unity.
	\par
	The total \gls{wsr} is given by $\rate_\Sigma =\sum_{i=1}^{I} \alpha_i \rate_i$, where $\alpha_i>0$ are predefined weights.
	Hence, the problem of maximizing the \gls{wsr} under maximum power constraints can be formulated as
	\vspace{-13pt}
	\begin{align}
		\max_{\matset{\matr{V}_i}_{i=1}^I} ~& \sum_{i=1}^{I} \alpha_i \logdet{\identity + \matr{Q}_i \matr{Z}_i \inv} \nonumber\\
		~ \subjectto ~&
		\sum_{i \in \mathcal{I}_k} \fronorm{\matr{V}_i}^2 \leq P_k,~ \forall k \in \matset{1, \dots, K}, \label{eq:main:main_sum_rate_problem}
	\end{align}
	\vspace{-3pt}
	where $P_k$ for $k=1,\dots, K$ are the \gls{bs} power budgets.
	In general, the objective function is highly multimodal,
	making a good optimum generally difficult to find \cite{lou2008_spectrum_management,hunger2008_combinatorial_approach_mimobc_nonconvex}.
	It must be remarked that although the above channel model is common in literature, the frequency flatness and instantaneous rate assumption are simplifications.
	
	\subsection{WMMSE Algorithm}
	The authors of \cite{shi2011_wmmse} tackle problem \eqref{eq:main:main_sum_rate_problem} by reformulating it to the \gls{wmmse} minimization
	\vspace{-2pt}
	\begin{align}
		\min_{\matset{\matr{U}_i, \matr{W}_i, \matr{V}_i}_{i=1}^I} ~&
		\sum\nolimits_{i=1}^{I} \alpha_i \left(\trace{\matr{W}_i \matr{E}_i} - \logdet{\matr{W}_i}\right) \nonumber\\
		\subjectto ~& 
		\sum\nolimits_{i \in \mathcal{I}_k} \fronorm{\vm_i}^2 \leq P_k ,~ \forall k \in \matset{1, \dots, K}\nonumber \\& \matr{W}_i \succeq\nullmat, ~\forall i \in \matset{1, \dots, I}, \label{eq:main:reform_sum_rate_problem}
	\end{align}\vspace{-14pt}\\
	where $\matr{E}_i = \expectation{}{(\breve{\vect{s}}_i - \matr{U}_i\herm\vect{y}_{\mr{Rx},i})(\breve{\vect{s}}_i - \matr{U}_i\herm\vect{y}_{\mr{Rx},i})\herm}$ is the error covariance of the estimated symbol vector using the linear receive beamformer $\matr{U}_i\in\Cset^{N_i\times N_i}$ and $\matr{W}_i$ is the \gls{psd} weight matrix assigned to \gls{ue} $i$.
	Problem \eqref{eq:main:main_sum_rate_problem} and \eqref{eq:main:reform_sum_rate_problem} share the optimal point and their stationary points.
	In a slight abuse of notations, $\matr{U}$, $\matr{W}$ and $\matr{V}$ are defined as short hand for their respective set of receive beamformers $\matset{\matr{U}_i}_{i=1}^I$, weight matrices $\matset{\matr{W}_i}_{i=1}^I$ and downlink beamformers $\matset{\matr{V}_i}_{i=1}^I$.
	It is straightforward to show that the \gls{wmmse} objective in \eqref{eq:main:reform_sum_rate_problem} is convex w.r.t. $\matr{U}$, $\matr{W}$ and $\matr{V}$ individually.
	By subsequently deriving their first-order stationary points, the \gls{bcd} algorithm \cite{bertsekas1999_nonlinear_programming} with updates
	\begin{subequations}
		\begin{align}
			(\matr{U}\text{-Step}) && \forall i:~& \matr{U}_i\itindex{\ell} = (\matr{J}_i\itindex{\ell})\inv\matr{H}_{ik}\matr{V}_i\itindex{\ell-1} \label{eq:theory:mmsebf}\\
			(\matr{W}\text{-Step}) && \forall i:~& \matr{W}_i\itindex{\ell} = \left(\identity - (\matr{V}_i\itindex{\ell-1})\herm \matr{H}_{ik}\herm \matr{U}_i\itindex{\ell}\right)\inv \label{eq:theory:wupdate}\\
			(\vect{\mu}\text{-Step}) && \forall k:~& \mu_k \itindex{\ell} = \argmax_{\mu_k} \mu_k \nonumber\\ &&&\text{subject to}~ 0= f_{\mr{CS},k}\itindex{l} (\mu_k), ~ \mu_k \geq 0, \label{eq:theory:dualfun}\\
			(\matr{V}\text{-Step}) && \forall i:~& \matr{V}_i\itindex{\ell} = \left(\matr{R}_k\itindex{\ell}  + \mu_k\itindex{\ell} \identity \right)^\dagger \vmtilde_i\itindex{\ell} \label{eq:theory:vupdate} %
		\end{align}
	\end{subequations}
	is obtained \cite{shi2011_wmmse}. Note that the indices $k$ and $i$ are chosen such that $i \in \setik$ throughout. 
	The updates are performed sequentially.
	Here we defined $\matr{J}_i\itindex{\ell} = \sum\nolimits_{m=1}^K \sum\nolimits_{j \in \mathcal{I}_m} \matr{H}_{i m} \matr{V}_j\itindex{\ell-1} (\matr{V}_j\itindex{\ell-1})\herm \matr{H}_{i m}\herm + \sigma_i^2\identity$ as the receive signal covariance matrix at \gls{ue} $i$, 
	\begin{equation}
			\setlength{\abovedisplayskip}{3pt}
			\setlength{\belowdisplayskip}{3pt}
			\matr{R}_k\itindex{\ell} = \sum\nolimits_{j=1}^{I} \alpha_j \matr{H}_{jk}\herm \matr{U}_j\itindex{\ell} \matr{W}_j\itindex{\ell} (\matr{U}_j\itindex{\ell})\herm \matr{H}_{jk} \label{eq:theory:weighteduplinkcov}
	\end{equation}
	as the weighted uplink covariance matrix at \gls{bs} $k$ and
	\begin{equation}
			\setlength{\abovedisplayskip}{3pt}
			\setlength{\belowdisplayskip}{3pt}
			\vmtilde_i\itindex{\ell} = \alpha_i \matr{H}_{ik}\herm \matr{U}_i\itindex{\ell} \matr{W}_i\itindex{\ell} \label{eq:theory:vcandidate}
	\end{equation} 
	as the candidate beamformer matrix of \gls{ue} $i \in \setik$ at \gls{bs} $k$ respectively.
	Note that $\matr{W}_i$ in \eqref{eq:theory:wupdate} is merely the inverse of the resulting \gls{mmse} covariance matrix $\matr{E}_i^\mr{MMSE}$ at the receiver.
	Equations \eqref{eq:theory:dualfun} and \eqref{eq:theory:vupdate} result from applying \gls{kkt} conditions to solve for $\matr{V}_i\itindex{\ell}$ under the constraint $\sum_{i \in \mathcal{I}_k} ||\matr{V}_i\itindex{\ell}||_F^2 \leq P_k$, thereby introducing dual variables $\mu_k\itindex{l}$ for ${k=1,\dots,K}$ at iteration $\ell$, as well as the complementary slackness condition $f_{\mr{CS},k}\itindex{l} (\mu_k) = 0$ where
	\begin{align}
		f_{\mr{CS},k}\itindex{\ell}(\mu_k) &= \left(\upsilon_k \itindex{\ell} (\mu_k) - 1\right)\mu_k \label{eq:theory:compslack}\\
		\text{and} \qquad \upsilon_k \itindex{\ell} (\mu_k ) &= \sum_{m=1}^{M_k} \frac{\varphi_{km}\itindex{\ell}}{(\lambda_{km}\itindex{\ell} + \mu_k)^2} \label{eq:theory:ratfun}.
	\end{align}
	The quantities $\lambda_{km}\itindex{\ell} = \matel{\matr{\Lambda}_k\itindex{\ell}}{mm}$ and $\varphi_{km}\itindex{\ell} = \matel{\matr{\Phi}_k\itindex{\ell}}{mm}$ are obtained from the eigendecomposition of the normal matrix $\matr{R}_k\itindex{\ell} = \matr{D}_k\itindex{\ell} \matr{\Lambda}_k\itindex{\ell} (\matr{D}_k\itindex{\ell})\herm$ and
	$
		\matr{\Phi}_k\itindex{\ell} = \frac{1}{P_k} (\matr{D}_k \itindex{\ell})\herm \left(\sum\nolimits_{i \in \mathcal{I}_k} \vmtilde_i\itindex{\ell} (\vmtilde_i\itindex{\ell})\herm \right) \matr{D}_k\itindex{\ell},
	$ respectively.
	To satisfy the complementary slackness condition \eqref{eq:theory:compslack}, the authors of \cite{shi2011_wmmse} propose the bisection search to find the root of $\upsilon_k (\mu_k) - 1 = 0$.
	Shi \etal establish convergence of the WMMSE algorithm to a stationary point. %
	It is critical to remark that, compared to \cite{shi2011_wmmse}, we apply the Moore-Penrose inverse in \eqref{eq:theory:vupdate} instead of the standard inverse.
	Update \eqref{eq:theory:vupdate} is the \textit{exact} minimizer of the block variable and covers particular occurring instances in which the matrix $\matr{R}_k\itindex{\ell} + \mu_k\itindex{\ell} \identity$ proves to be singular and the original update rule does not apply.
	\par
	Importantly, the algorithm can be implemented in a distributed fashion by computing $\matr{U}$-updates and $\matr{W}$-updates at the respective \glspl{ue} and each \gls{bs} computes $\matr{V}$ only for its assigned beamformers.
	Each \gls{ue} needs to locally estimate $\matr{J}_i$ and requires information about their assigned precoding matrix as well as the channel toward its assigned \gls{bs}.
	The \glspl{bs} on the other hand require local CSI, the matrix $\matr{U}_i\matr{W}_i\matr{U}_i\herm$ for each \gls{ue} within their cell radius as well as the matrix $\um_i \wm_i$ of their assigned \glspl{ue}.
	\section{Proposed WMMSE Unfolding Architecture} \label{sec:model}
	\begin{figure*}[t]
		\setlength{\abovecaptionskip}{-6pt}
		\centering
		\def\svgwidth{\linewidth}
		\input{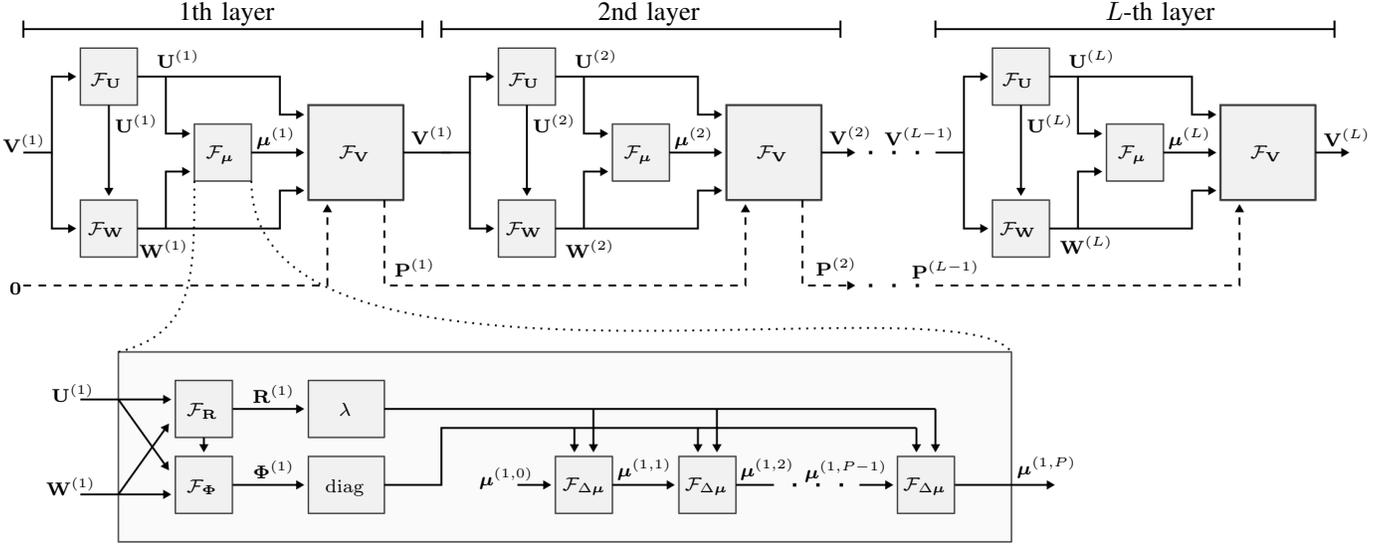}
		\caption{Deep network of the WMMSE algorithm obtained by unrolling $L$ iterations. The blocks $\mathcal{F}_{\matr{U}}$, $\mathcal{F}_{\matr{W}}$ and $\mathcal{F}_{\matr{V}}$ represent the updates \eqref{eq:theory:mmsebf}, \eqref{eq:theory:wupdate} and \eqref{eq:theory:vupdate} for every $i$. The $\mathcal{F}_{\vect{\mu}}$-block contains ${P}$ root-finding iterations $\mathcal{F}_{\Delta \vect{\mu}}$ representing \eqref{eq:main:rootiter} for every $k$. The blocks $\lambda$ and $\operatorname{diag}$ are the extraction of eigenvalues and diagonals respectively. The GCN-WMMSE architecture modifies the $\mathcal{F}_{\matr{W}}$- and $\mathcal{F}_{\matr{V}}$-blocks of the original algorithm. Dashed arrows represent the skip connections proposed in Section \ref{sec:main:layerconnection}.}
		\label{fig:main:gcnwmmse_layer}
		\vspace{-12pt}
	\end{figure*}
	In this section, we introduce the proposed \textbf{G}raph-\textbf{C}onvolutional-\textbf{N}etwork-WMMSE architecture, or GCN-WMMSE in short.
	First, in Section \ref{sec:main:differentiability}, we review the concept of unfolding and define a neural network based on the general \gls{wmmse} as our starting point. %
	In Section \ref{sec:main:architecture}, the GCN-WMMSE architecture is obtained by unfolding the baseline WMMSE algorithm. 
	It aims to reduce the required number of iterations, leading to a lower computational cost and overhead in distributed deployments.
	Section \ref{sec:main:training} details the training procedure that will be used to learn the parameters of GCN-WMMSE networks.
	In Section \ref{sec:main:model_analysis}, the proposed architecture is analyzed.

	\subsection{WMMSE Network} \label{sec:main:differentiability}
	\par
	Let $\mathcal{P}=(\matset{M_k, P_k, \setik}_{k=1}^K, \matset{N_i, \sigma_i^2, \alpha_i}_{i=1}^I, \mathbb{H})$ be a tuple of wireless scenario parameters defining a network topology and channel conditions where $\mathbb{H}$ describes a distribution of conforming channel matrix sets $\matset{\hmat_{ik}}_{i,k}$. %
	Let $\mathcal{S} = (\mathcal{P}, \matset{\hmat_{ik}}_{i,k})$ be a wireless scenario \textit{realization} containing sampled channel matrices based on some scenario configuration $\mathcal{P}$, and let the beamformer achieving the maximum \gls{wsr} be ${{}^{\mathcal{S}}\vm\itindex{\mr{opt}}= \argmax_\vm \rate_\Sigma (\vm; \mathcal{S})}$.
	We seek a data-driven neural network model $\mathcal{M}(\mathcal{S}; \vect{\Gamma})$ parameterized by a set of network parameters $\vect{\Gamma}$ that approximates ${}^{\mathcal{S}}\vm\itindex{\mr{opt}}$ such that the expected rate over the distribution of scenarios is similar: %
	\begin{equation}
		\expectation{\mathcal{S} \sim \prob{\mathcal{S}}}{ \rate_\Sigma ({}^{\mathcal{S}}\vm\itindex{\mr{opt}}; \mathcal{S})} \approx \expectation{\mathcal{S} \sim \prob{\mathcal{S}}}{ \rate_\Sigma (\mathcal{M}(\mathcal{S}; \vect{\Gamma}); \mathcal{S})}. \label{eq:main:goal}
	\end{equation}
	The concept of unfolding entails adopting the architecture of a neural network from a hand-crafted iterative algorithm, then modifying and parameterizing it.
	By learning the optimal value of the network parameters from data, a network that is at least as performant as the original method while being computationally more efficient can be obtained \cite{monga2020_algorithm_unrolling, balatsoukasstimming2019_deepunrolling_comm}. 
	\par
	We base the architecture of GCN-WMMSE on $L$ iterations of the \gls{wmmse} algorithm by interpreting each iteration with input $\vm\itindex{\ell-1}$ and output $\vm\itindex{\ell}$ as a neural network layer $\mathcal{M}_\mr{u}\itindex{\ell}(\vm; \mathcal{S})$, such that the full network is the composition
	$
		\mathcal{M}_\mr{u}(\vm; \mathcal{S}) = (\mathcal{M}_\mr{u}\itindex{L} \circ \mathcal{M}_\mr{u}\itindex{L-1} \circ \dots \circ \mathcal{M}_\mr{u}\itindex{1})(\vm; \mathcal{S})
	$ \wrt $\vm$.
	Figure \ref{fig:main:gcnwmmse_layer} visualizes the forward pass of the resulting network. %
	Each layer $\mathcal{M}_\mr{u}\itindex{\ell}$ consists of the update blocks $\mathcal{F}_{\wm}(\vm; \mathcal{S})$, $\mathcal{F}_{\um}(\vm, \wm; \mathcal{S})$ and $\mathcal{F}_{\vm}(\um, \wm, \vect{\mu}; \mathcal{S})$, which are the update mappings \eqref{eq:theory:mmsebf}, \eqref{eq:theory:wupdate} and \eqref{eq:theory:vupdate} for every \gls{ue} $i$, %
	and $\mathcal{F}_{\vect{\mu}}(\um, \wm; \mathcal{S})$ which is \eqref{eq:theory:dualfun} for every \gls{bs} $k$.
	\par
	Previous works \cite{chowdhury2021_unfolding, pellaco2020_deep_unfolding_pgd, hu2021_iadnn, chowdhury2021_mlaided_tacticalmimo} are restricted to less general wireless scenarios in order to unroll a simplified \gls{wmmse} network.
	For example, the utilization of \gls{kkt} conditions is avoided by restricting to single-cell networks \cite{hu2021_iadnn}, or the optimization variables is made real and scalar by restricting to pairwise \gls{siso} links \cite{chowdhury2021_unfolding}.
	In our proposed network architecture in comparison, the general WMMSE algorithm is considered.
	Unfortunately, in this case the mapping $\mathcal{F}_{\vect{\mu}}$, \ie, the solution to the problem in \eqref{eq:theory:dualfun}, does not possess a closed-form solution in general.
	The iterative bisection search proposed in \cite{shi2011_wmmse} to compute the optimum $\mu_k\itindex{\ell, \mr{opt}}$ is inadvisable since its accuracy is insufficient for a low number of subiterations ${P}$ in the forward pass of the network.
	Instead, we propose to utilize a method based on rational function approximations \cite{liu2022_successive_ratfunproof} to accelerate the forward pass of the \gls{wmmse} network and algorithm.
	At each substep ${p}$ of the root-finding algorithm, an approximation $\tilde{\upsilon}(\mu_k\itindex{\ell,{p}}, \delta_\mr{N}, \delta_\mr{D}) =  \delta_\mr{N} / (\delta_\mr{D}+\mu_k\itindex{\ell,{p}})^2$ of \eqref{eq:theory:ratfun} at the current iterate $\mu_k\itindex{\ell,{p}}$ is computed, where $\tau$ and $\delta$ are determined such that $\tilde{\upsilon}(\mu_k\itindex{\ell,{p}}, \delta_\mr{N}, \delta_\mr{D}) = \upsilon_k\itindex{\ell}(\mu_k\itindex{\ell,{p}})$ and $\tilde{\upsilon}'(\mu_k\itindex{\ell,{p}}, \delta_\mr{N}, \delta_\mr{D}) = {\upsilon_k'}\itindex{\ell}(\mu_k\itindex{\ell,{p}})$.
	The non-negative intersection of $\tilde{\upsilon}$ with $1$ can be obtained in closed-form, leading to the update
	\begin{equation}
		\setlength{\abovedisplayskip}{0pt}
		\setlength{\belowdisplayskip}{4pt}
		\mu_k\itindex{\ell,{p}+1} = \mu_k\itindex{\ell,{p}} + 2\frac{\upsilon_k\itindex{\ell}(\mu_k\itindex{\ell,{p}})}{{\upsilon_k'}\itindex{\ell}(\mu_k\itindex{\ell,{p}})}\left(1 - \sqrt{\upsilon_k\itindex{\ell}(\mu_k\itindex{\ell,{p}})}\right) \label{eq:main:rootiter}
	\end{equation}
	which can be shown to converge monotonically to the root $\mu_k\itindex{\ell,\mr{opt}}$ if $0\leq\mu_k\itindex{\ell,{p}} \leq \mu_k\itindex{\ell,\mr{opt}}$.
	If $\mu_k\itindex{\ell,\mr{opt}} < \mu_k\itindex{\ell,{p}}$, the iterate jumps to the interval $(-\infty, \mu_k\itindex{\ell,\mr{opt}})$ within a single iteration.
	Our simulations show that floating point precision is achieved after a low number of substeps ${P}$.
	To extend the solution to solve the problem in \eqref{eq:theory:dualfun}, $\mu\itindex{\ell, {p}}$ is further mapped to $0$ after each iteration if it is negative.
	The substeps \eqref{eq:main:rootiter} for every $k$ are depicted as $\mathcal{F}_{\Delta \vect{\mu}}$-blocks in \fig{fig:main:gcnwmmse_layer}.
	\subsection{GCN-WMMSE Architecture} \label{sec:main:architecture}
	\par
	Depending on the scenario realization $\mathcal{S}$, the \gls{wmmse} network requires many layers to approach a local optimum.
	In this work we modify the $\mathcal{F}_\wm$-blocks and $\mathcal{F}_\vm$-blocks using concepts from \glspl{gcf} and \glspl{gcn} to significantly reduce the number of required layers while achieving beamformers of comparable quality. %
	To this end, we first briefly introduce \glspl{gcf} and \glspl{gcn} in the following.
	\subsubsection{Graph Convolutional Filters/Networks}
	\par
	A \gls{gcf} of order $G$ is given as
	\begin{equation}
		\setlength{\abovedisplayskip}{1pt}
		\setlength{\belowdisplayskip}{5pt}
		\vect{f}_\mr{GCF}(\matr{x}; \matr{S}, \matset{a_g}_{g=0}^G) = \sum\nolimits_{g=0}^G a_g \matr{S}^g \vect{x}, \label{eq:main:gcf}
	\end{equation}
	where $a_g$ for $g=0,\dots,G$ are filter taps.
	The matrix $\matr{S}\in\Rset^{M \times M}$ is a shift matrix which describes the progression of a graph signal $\vect{x}\in \Rset^M$ between two discrete time steps.
	\par
	\glspl{gnn} are an extension to graph filters which enable a multilayer design by introducing nonlinearities with multiple parallel filters similar to convolutional neural networks.
	A single layer of a \gls{gcn}, a common variant of a \gls{gnn}, can be expressed as
	\begin{equation}
		\setlength{\abovedisplayskip}{0pt}
		\matr{F}_\mr{GCN}(\matr{X}; \matr{S}, \matset{\matr{A}_g}_{g=0}^G, \vect{b}) = \phi\left(\sum_{g=0}^G \matr{S}^g\matr{X}\matr{A}_g + \onevec\matr{b}\tran\right). \label{eq:main:gcn_layer}
	\end{equation}
	Equation \eqref{eq:main:gcn_layer} maps a signal $\matr{X}\in\Rset^{N\times F_1}$ containing $F_1$ features to an output of size ${N\times F_2}$ with $F_2$ features.
	The matrices $\matr{A}_g \in \Rset^{F_1 \times F_2}$ for $g=0,\dots,G$ store the filter coefficients.  
	In conjunction with a biasing parameter $\vect{b}\in \Rset^F$, an elementwise nonlinearity $\phi$ furthers expressiveness of the mapping, allowing for higher selectivity by frequency mixing while maintaining stability to errors \cite{gama2019_stability}.
	Please refer to \cite{shuman2012_gsp, gama2019_gnn_architectures} for more details.
	In the following, we unfold the $\mathcal{F}_\wm$- and $\mathcal{F}_\vm$-blocks.
	
	\subsubsection{Weight Matrix Graph Filter}
	\par
	The idea of graph filtering is extended to the complex domain and applied to the $\mathcal{F}_\wm$-block.
	To this end, we consider the original weight matrix update in \eqref{eq:theory:wupdate} ${\hat{\wm}\itindex{\ell} = \left(\identity - (\matr{V}_i\itindex{\ell-1})\herm \matr{H}_{ik}\herm \um\itindex{\ell} \right)\inv}$, where $i \in \setik$, as a shift matrix and introduce the weight matrix \gls{gcf}
	\begin{align}
		\matr{W}_i\itindex{\ell} = a_{\mr{W},\ell 0} \identity + \sum_{g=1}^{G} \frac{a_{\mr{W},\ell g}}{\left(\trace{\hat{\matr{W}}_i\itindex{\ell}} / N_i\right)^{g-1}}\left(\hat{\matr{W}}_i\itindex{\ell}\right)^g \label{eq:main:weightgcf}
	\end{align}
	of order $G$ for every  $i$, where $a_{\mr{W},\ell g}$ for $g=0,\dots,G$ and $\ell=1,\dots,L$, are learnable filter taps, %
	Thus, \eqref{eq:main:weightgcf} replaces the standard \gls{wmmse} weight update. %
	The filter taps $a_{\mr{W},\ell g}$ are restricted to be non-negative ($a_{\mr{W},\ell g} \geq 0$) to ensure that $\matr{W}_i\itindex{\ell}$ is \gls{psd}.
	The mean of eigenvalues $\trace{\hat{\wm}_i\itindex{\ell}}/N_i$ normalizes the filter taps for $g\geq2$ to prevent numerical issues in case of a high \gls{snr}. %
	It can be shown that an equal scaling transformation of $\matr{W}_i\itindex{\ell}$ for all $i$ does not change the output of a layer $\ell$ if the following update blocks follow the normal \gls{wmmse} updates.
	However, this scaling ambiguity can be prevented by fixing one filter tap $a_{\mr{W},\ell g}$, or a modification of the subsequent $\mathcal{F}_\vm$-block as outlined below.
	
	\subsubsection{Downlink Graph Convolutional Neural Network}
	\par
	The $\matr{V}$-step in \eqref{eq:theory:vupdate} can be interpreted as a \gls{gcf} with the pseudoinverse of the modified weighted uplink covariance matrix ${\tilde{\matr{R}}_k\itindex{\ell} = \matr{R}_k\itindex{\ell} + \mu_k\itindex{\ell,{P}}\identity}$ in \eqref{eq:theory:weighteduplinkcov} as shift operator.
	The filter acting on the candidate beamformer $\vmtilde_i\itindex{\ell}$ in \eqref{eq:theory:vcandidate} is then extended into a complex-valued \gls{gcn} layer with $F$ features as in \eqref{eq:main:gcn_layer}, leading to the modified $\matr{V}$-update obtaining the unscaled beamformer $\hat{\vm}_i$
	\begin{align}
		\hat{\vect{v}}_{id}\itindex{\ell} &= \modrelu\left(\tilde{\matr{P}}_{id}\itindex{\ell}, \frac{1}{b_\mr{S}}  \sqrt{\frac{P_k}{\abs{\mathcal{I}_k}}}\onevec \vect{b}_{\ell}\tran\right) \vect{c}_\ell\nonumber\\
		\text{where}\quad & \tilde{\matr{P}}_{id}\itindex{\ell} = (\tilde{\matr{R}}_k\itindex{\ell})^\dagger \vvtilde_{id}\itindex{\ell} \vect{a}_{\mr{V},\ell 1}\tran + \vvtilde_{id}\itindex{\ell} \vect{a}_{\mr{V},\ell 0}\tran \label{eq:main:gcnupdate} %
	\end{align}
	for every $i\in \setik$ for all $k$, where $\vvtilde_{id}\itindex{\ell}$ and $\hat{\vv}_{id}\itindex{\ell}$ for $d=1,\dots, N_i$ are the columns/streams in $\vmtilde_i\itindex{\ell} = \vecbrac{\vvtilde_{i1}\itindex{\ell},\dots, \vvtilde_{i N_i}\itindex{\ell}}$ and $\hat{\vm}_i\itindex{\ell} = \vecbrac{\hat{\vv}_{i1}\itindex{\ell},\dots, \hat{\vv}_{i N_i}\itindex{\ell}}$.
	The \gls{gcn} layer corresponds to a filter of polynomial degree $1$ with $\vect{a}_{\mr{V},\ell 1}\in\Cset^F$ and $\vect{a}_{\mr{V},\ell 0}\in\Cset^F$ being trainable complex filter taps,  $\vect{b}_\ell\in\Rset^F$ being an additional trainable bias, and $\vect{c}_\ell \in \Cset^F$ is a trainable vector recombining the $F$ features contained in $\tilde{\matr{P}}_{id}\itindex{\ell}$.
	Higher polynomial orders empirically do not provide any benefits.
	The bias is scaled by the power and number of assigned \glspl{ue} in order to improve the generalization performance.
	The auxiliary scaling parameter $b_\mr{S}$ reduces the magnitude of the bias term relative to the other summands, while maintaining the flexibility for different numbers of assigned \glspl{ue} and power changes.
	This becomes necessary when applying an optimizer like ADAMW \cite{loshchilov2019_decoupled_weight_decay} to prevent bias parameters steps having a too dominant impact compared to other parameters.
	$\modrelu$ is a complex variant of the rectified linear unit ($\relu$) \cite{trabelsi2017_deep_complex_networks} and is defined as
	\begin{align}
		\modrelu(x, b) = \begin{cases}
			(\abs{x} + b) \frac{x}{\abs{x}}, & \text{for~}\abs{x} + b > 0\\
			0, & \text{otherwise.}
		\end{cases}\label{eq:main:modrelu}
	\end{align}
	The nonlinearity \eqref{eq:main:modrelu} empirically performs better than applying a $\relu$ to the real and imaginary components individually.
	The architecture is limited to single-layer \glspl{gcn} since the benefit of multiple layers has been contested \cite{wu2019_simple_gcn}. 
	Furthermore, applying multi-layer \glspl{gcn} does not yield an empirical improvement for GCN-WMMSE. 
	Since adherence to the power constraints cannot be guaranteed by the \gls{gcn}, an additional power projection step
	\begin{align}
		\matr{V}_i\itindex{\ell} = \sqrt{P_k}\hat{\matr{V}}_i\itindex{\ell} / \sqrt{\max\left\{\sum_{i\in\mathcal{I}_k}\fronorm{\hat{\matr{V}}_i\itindex{\ell}}^2, P_k\right\}},
		\label{eq:main:pownorm}
	\end{align}
	for all $i$ is introduced as a post-processing step of the output of each $\mathcal{F}_\vm$-block.
	Sets of beamforming matrices $\{\hat{\vm}_i\itindex{\ell}\}_{i\in \setik}^{I}$ exceeding the maximum power $P_k$ are projected onto the feasible set, while feasible beamformer sets are preserved.
	
	\subsubsection{Skip Connections} \label{sec:main:layerconnection}
	\par
	Lastly, we adopt the concept of additional connections between layers that has been successfully applied in residual and highway networks \cite{he2015_resnet,srivastava15_highwaynetworks,rahimi2018_gnn_geolocation_highway}.
	Specifically, the input to the $\modrelu$ nonlinearity $\tilde{\matr{P}}_{id}\itindex{\ell}$ is replaced by $\matr{P}_{id}\itindex{\ell} = \tilde{\matr{P}}_{id}\itindex{\ell} + \matr{P}_{id}\itindex{\ell-1}\matr{D}_{\ell}$.
	A trainable mapping matrix $\matr{D}_{\ell}\in\Cset^{F \times F}$ allows for linear combinations of the features of the previous layer.
	Thus, a direct path for the additional exchange of gradient information between layers is created, bypassing operations such as matrix inversions which sometimes lead to noisy gradients.
	The additional network connections are illustrated in \fig{fig:main:gcnwmmse_layer} as dashed arrows.
	\subsubsection{Initialization Beamformers}
	The function generating the initial beamformer set $\vm\itindex{0}$ that is fed into the input of the network needs to be integrated into the network models themselves since the optimal values of the trainable parameters depend on it.
	In this work we only consider initialization functions which are entirely determined by the scenario realization $\mathcal{S}$ without any additional trainable parameters.
	We thus obtain deep networks $\mathcal{M}(\mathcal{S}; \vect{\Gamma})$ with a trainable parameter set $\vect{\Gamma}$ containing in total $\{\vect{a}_{\mr{V},\ell 1}\}_{\ell=1}^L$, $\{\vect{a}_{\mr{V},\ell 0}\}_{\ell=1}^L$, $\{\vect{b}_\ell\}_{\ell=1}^L$, $\{\vect{c}_\ell\}_{\ell=1}^L$, $\{\vect{D}_\ell\}_{\ell=2}^L$ and $\{\{a_{\mr{W},\ell g}\}_{g=0}^G\}_{\ell=1}^L$ in case of $L$ layers
	($b_\mr{S}$ is separately treated as discussed in Section \ref{sec:main:training}).
	
	\subsection{Model Training} \label{sec:main:training}
	Optimizing the trainable parameter set $\vect{\Gamma}$ of a GCN-WMMSE network with $L$ layers is achieved by the maximization of the expected rate in \eqref{eq:main:goal}. %
	In practice, the expectation needs to be approximated by a finite data set $\mathcal{T}$ and the non-convex optimization is classically performed by \gls{sgd}.
	In this work, we perform \gls{sgd} for $T$ steps by descending along the gradient $\nabla_{\matr{\Gamma}}J(\mathcal{T}_t; \matr{\Gamma}, \mathcal{L})$ of the sample-normalized loss function
	\begin{equation}
		\setlength{\abovedisplayskip}{0pt}
		J(\mathcal{T}_t; \matr{\Gamma}, \mathcal{L}) = \frac{1}{\abs{\mathcal{T}_t}\abs{\mathcal{L}}} \sum_{\mathcal{S}_n\in\mathcal{T}_t}\sum_{\ell\in\mathcal{L}} \frac{ J_\mr{WSR}\itindex{\ell}(\mathcal{S}_n;\matr{\Gamma})}{r_{n, \matr{\Gamma}}\itindex{\ell}}, \label{eq:main:normloss}
	\end{equation}
	where $\mathcal{L}$ is a set of layers and $\mathcal{T}_t$ is a minibatch of scenario realizations $\mathcal{S}_n$ at training step $t$.
	The partial loss $J_\mr{WSR}\itindex{\ell}$ is the negative \gls{wsr} achieved on the realization $\mathcal{S}_n$ with the downlink beamformer set $\matr{V}\itindex{\ell}$ computed by the $\ell$-th network layer given a set of network parameters $\matr{\Gamma}$, \ie, $J_\mr{WSR}\itindex{\ell}(\mathcal{S}_n;\matr{\Gamma}) = -\mathcal{R}_\Sigma (\mathcal{M}\itindex{1:\ell}(\mathcal{S}_n; \vect{\Gamma});\mathcal{S}_n)$
	where $\mathcal{M}\itindex{1:\ell}$ denotes the neural network up to layer $\ell$.
	The scalar $r_{n, \vect{\Gamma}}\itindex{\ell}$ is determined as the magnitude of $J_\mr{WSR}\itindex{\ell}(\mathcal{S}_n;\matr{\Gamma})$ in the forward pass.
	It equalizes the loss per sample to unity but is otherwise treated as scaling in the gradient computation.
	Thus, realizations with lower achievable \gls{wsr} have the same impact on the gradient as realizations with high \gls{wsr}.
	Equation \eqref{eq:main:normloss} generalizes the loss function utilized in \cite{alkhateeb2018_deeplearned_coordinatedbf} for multiple output layers.
	The scaling parameter $b_\mr{S}$ in \eqref{eq:main:gcnupdate} is obtained during training by a running mean of the empirical expectation $\expectation{\mathcal{S} \sim \mathcal{T}}{\sqrt{\frac{P_k}{\abs{\mathcal{I}_k}}}}$ dependent on the power budget of scenarios in the training set $\mathcal{T}$.
	\par
	\gls{sgd} involves the full complex gradient \wrt the parameters in $\vect{\Gamma}$ which is typically obtained by backpropagation.
	Since this requires the full Jacobian \wrt the inputs of the update blocks, \eg, $\vm$ and $\wm$ in case of the $\mathcal{F}_{\wm}$-block, the existence of the partial derivatives must be ensured.
	This is straightforward for all matrix operations in the GCN-WMMSE architecture, and workarounds for pointwise non-differentiable nonlinearities are well investigated.
	However, naive backpropagation of multiple concatenated update steps \eqref{eq:main:rootiter} in the $\mathcal{F}_{\vect{\mu}}$-block to obtain the derivative \wrt the eigenvalues of $\rmat_k\itindex{\ell}$ and the diagonal values $\varphi_{km}\itindex{\ell}$, while feasible in practice, is prone to numerical issues as it involves divisions by small numbers. %
	Instead, an efficient and exact gradient of the entire update block $\mathcal{F}_{\vect{\mu}}$ assuming a fully converged iterate can be obtained within a single step. 
	\newtheorem{prop}{Proposition}
	\begin{prop} \label{prop:complementary_derivative}
		Let $\mu_k\itindex{\ell, \mr{opt}} = f(\vect{z}_k\itindex{\ell})$ where ${\vect{z}_k\itindex{\ell} = \left(\matset{\varphi_{km}\itindex{\ell}}_{m=1}^M, \matset{\lambda_{km}\itindex{\ell}}_{m=1}^M\right)}$ be the function that is implicitly defined as the solution of \eqref{eq:theory:dualfun}. 
		For any point $\tilde{\vect{z}}_k\itindex{\ell}$ with  $\tilde{\varphi}_{km}\itindex{\ell} > 0$ and $\tilde{\lambda}_{km}\itindex{\ell} > -\tilde{\mu}_k\itindex{\ell, \mr{opt}}$: The gradient of $f$ at $\tilde{\vect{z}}_k\itindex{\ell}$ exists if $\tilde{\mu}_k\itindex{\ell, \mr{opt}} >0$ and
		\begin{align}
			\frac{\partial}{\partial \varphi_{ki}\itindex{\ell}} f(\tilde{\vect{z}}_k\itindex{\ell}) =  \frac{(\tilde{\lambda}_{ki}\itindex{\ell} + \tilde{\mu}_k\itindex{\ell, \mr{opt}})^{-2}}{2  \sum_{m=1}^{M_k} \frac{\tilde{\varphi}_{km}}{(\tilde{\lambda}_{km} + \tilde{\mu}_k\itindex{\ell, \mr{opt}})^3}},\nonumber\\
			\frac{\partial}{\partial \lambda_{ki}\itindex{\ell}} f(\tilde{\vect{z}}_k\itindex{\ell}) = -\frac{\tilde{\varphi}_{ki}\itindex{\ell} (\tilde{\lambda}_{ki}\itindex{\ell} + \tilde{\mu}_k\itindex{\ell, \mr{opt}})^{-3}}{ \sum_{m=1}^{M_k} \frac{\tilde{\varphi}_{km}}{(\tilde{\lambda}_{km} + \tilde{\mu}_k\itindex{\ell, \mr{opt}})^3}}.
		\end{align}
		Furthermore, for $\tilde{\mu}_k\itindex{\ell, \mr{opt}} =0$ with $\sum_{m=1}^{M} \frac{\tilde{\varphi}_{km}}{(\tilde{\lambda}_{km} + \tilde{\mu}_k\itindex{\ell, \mr{opt}})^2} \neq 1$, the gradient exists and is given by
		\begin{align}
			\frac{\partial}{\partial \varphi_{ki}\itindex{\ell}} f(\tilde{\vect{z}}_k\itindex{\ell})=0 \qquad \text{and} \qquad
			\frac{\partial}{\partial \lambda_{ki}\itindex{\ell}} f(\tilde{\vect{z}}_k\itindex{\ell})= 0.
		\end{align}
	\end{prop}
	A proof is obtained by applying \cite[Thm. 8.2]{amann2008_analysis2} to the complementary slackness condition \eqref{eq:theory:compslack}. %
	Since the derivative is undefined in the point where $\mu\itindex{\ell, \mr{opt}} = 0$ with $\sum_{m=1}^{M_k} \frac{\tilde{\varphi}_{km}}{(\tilde{\lambda}_{km} + \tilde{\mu}_k\itindex{\ell, \mr{opt}})^2} = 1$, it has to be extended in practice.
	In our implementation, it is set to $0$ in this case. %
	Furthermore, the exact solution $\tilde{\mu}_k\itindex{\ell, \mr{opt}}$ has to be replaced by the iterate $\tilde{\mu}_k\itindex{\ell, {P}}$.
	Note that a closed-form expression only exists for the derivative of the implicit function $f$. 
	In the forward pass of the network, a closed-form expression of $f$ does not exist and the iterative procedure with updates in \eqref{eq:main:rootiter} is used instead.
	
	\subsection{Discussion} \label{sec:main:model_analysis}
	Unfolding leads to a highly structured neural network with usually a low number of trainable parameters.
	In the case of the GCN-WMMSE architecture, the network models only contain $L(4F+G+1) + (L-1)F^2$ trainable parameters, which reduces the risk of overfitting and dramatically increases the data efficiency, as will be demonstrated in Section \ref{sec:experiments}. %
	Furthermore, it can be shown that the original \gls{wmmse} is contained in the family of functions spanned by the GCN-WMMSE architecture.
	\par
	Since GCN-WMMSE does not introduce additional quantities that need to be shared between consecutive update operations, with the exception of $\matr{P}_{id}\itindex{\ell}$ which remains local at a \gls{bs}, no additional information needs to be exchanged between \glspl{ue} and \glspl{bs} at test time compared to the \gls{wmmse}.
	Thus, assuming that the model parameters in $\vect{\Gamma}$ are already known throughout the network, GCN-WMMSE inherits the capability of distributed deployment from the WMMSE algorithm.
	In this scenario, \glspl{bs} only need to share $G+1$ parameters with the \glspl{ue} beforehand.
	\par
	The application of \glspl{gcf} and \glspl{gcn} to unfold the \gls{wmmse} algorithm inherits multiple favorable properties of those structures.
	Firstly, \glspl{gcf} and \glspl{gcn} exhibit permutation equivariance: Consider an arbitrary permutation matrix $\matr{\Pi}\in \matset{0,1}^{N \times N}$ fulfilling $\matr{\Pi}\onevec = \matr{\Pi}\tran\onevec = \onevec$ and a \gls{gcf} as in \eqref{eq:main:gcf}, then $\vect{f}_\mr{GCF}(\matr{x}; \matr{S}, \matset{a_g}_{g=0}^G) = \matr{\Pi}\vect{f}_\mr{GCF}(\matr{\Pi}\tran\matr{x}; \matr{\Pi}\tran\matr{S}\matr{\Pi}, \matset{a_g}_{g=0}^G)$. 
	The same holds for \gls{gcn} layers such as \eqref{eq:main:gcn_layer}.
	Consequently, relabeling of the nodes of a graph, which is equivalent to a permutation of the signal vector and equal permutation of the rows and columns of the shift matrix, does not change the output apart from said permutation.
	Permutation equivariance \wrt relabeling of the transceiver antennas is a natural property of a wireless network if no additional assumptions are made, and it is replicated by the original matrix filter operations of the \gls{wmmse} algorithm.
	Replacing those with \glspl{gcf} or \glspl{gcn} preserves this natural property in the GCN-WMMSE architecture; the proof is straightforward but is omitted for lack of space.
	Permutation equivariance \wrt relabeling of the network participants as emphasized in \cite{chowdhury2021_unfolding} is automatically achieved if all \glspl{bs} and \glspl{ue} share the same neural network parameters.
	\par
	Secondly, the filter matrix resulting from the \glspl{gcf} polynomial on the shift matrix $\matr{S}$ as in \eqref{eq:main:gcf} has the same eigenbasis as $\matr{S}$, but its eigenvalues are mapped by the polynomial with coefficients $\matset{a_g}_{g=0}^G$. %
	In case of the weight matrix update $\mathcal{F}_\wm$ \eqref{eq:main:weightgcf}, the coefficients of the polynom in $\hat{\wm}_i\itindex{\ell}$, which is the inverse of the receiver symbol error matrix $\matr{E}_i\itindex{l}$, are non-negative.
	Considering a degree $G \geq 2$, this leads to an amplification of large eigenvalues corresponding to low error eigenspaces and an attenuation of small eigenvalues corresponding to high error eigenspaces compared to a polynomial of degree $G=1$.
	Thus, when forming the candidate beamformer matrix $\vmtilde_i$ in \eqref{eq:theory:vcandidate} using $\wm_i\itindex{\ell}$, the \gls{gcf} acts as a highpass on the eigenvalues and corresponding signal spaces of the virtual uplink precoder $\alpha_i (\matr{U}_i\itindex{\ell})\herm \matr{H}_{ik}$ which belong to low error receive symbol spaces in $\matr{E}_i\itindex{l}$.
	Similarly, the significant interference eigenspaces of the weighted uplink covariance $\rmat_k\itindex{\ell}$ \eqref{eq:theory:weighteduplinkcov} are amplified due to the weight matrix \gls{gcf} as well.
	In the downlink beamformer update $\mathcal{F}_\vm$, the signal components of the candidate beamformer $\vmtilde_i$ are then filtered according to the eigenvalues of $\tilde{\rmat}_k\itindex{\ell}$.
	The property is known as interference avoidance and it is performed by the WMMSE algorithm as well as the proposed architecture both at the transmitter on the basis of the reciprocal channel and at the receiver.
	This scheme originates from cooperative approaches with the goal of interference alignment \cite{schmidt2009_wmmse_miso, gomadam2008_maxsinr}, which is optimal for high SNR.
	The WMMSE algorithm is similarly structured compared to algorithms which relax this goal \cite{peters2011_cooperative_algorithms}.
	The downlink beamformer update $\mathcal{F}_\vm$ in \eqref{eq:main:gcnupdate} is a trainable interference avoidance filter which directly leverages this notion.
	The \gls{gcn} provides additional degrees of freedom compared to a \gls{gcf} and simultaneously offers a higher selectivity due to the nonlinearity as well as resilience to errors \cite{gama2019_stability}.
	Both \eqref{eq:main:weightgcf} and \eqref{eq:main:gcnupdate} are optimized \wrt the global objective compared to the locally optimal operations \eqref{eq:theory:wupdate} and \eqref{eq:theory:vupdate} invoked by \gls{bcd}, thus achieving interference avoidance that is superior for the global \gls{wsr} objective. 
	\par
	Lastly, the applied \glspl{gcf} and \glspl{gcn} enable a high transferability regarding wireless scenario configurations.
	GCN-WMMSE networks are transferable to \textit{any} set of scenario configurations for any given set of trainable parameters.
	Therefore, it is more general than all previous works on unfolding the \gls{wmmse}.
	This, together with the parameter efficiency, can be leveraged in distributed training schemes such as federated learning.
	The generality of the model addresses the fundamental problem of heterogeneous data in federated learning and the low number of parameters increases the efficiency when exchanging model parameters or gradients.
	\par
	The proposed GCN-WMMSE architecture adds $\complexityupper{I(G-1)N^3 + IF^2NM + INM}$ of per-iteration complexity over the classical WMMSE algorithm, where $N$ and $M$ are the number of antennas of each \gls{ue} or \gls{bs}, respectively.
	However, the increase is absorbed into the WMMSE per-iteration complexity $\complexityupper{I^2N^2M + KIN^2M + INM^2 + IN^3 + K M^3}$, since $F$ and $G$ are constant \wrt the scenario.
	Thus, the impact of the additional operations is low, which is verified by experiments.

	\section{Experiments and Discussion} \label{sec:experiments}
	In this section, we evaluate the performance of the proposed GCN-WMMSE architecture with special focus on carving out its generalization capabilities, then we verify the application of those characteristics.
	All models and algorithms are implemented\footnote{To promote reproducible research, the code is publicly available at https://github.com/lsky96/gcnwmmse.} using PyTorch \cite{pytorch}.
	We apply \gls{sgd} and leverage the AdamW optimizer \cite{loshchilov2019_decoupled_weight_decay}.
	In Section \ref{sec:main:generalperf} and \ref{sec:main:generalization}, GCN-WMMSE network models are assessed on an artificial scenario of 3 \glspl{bs} positioned at the corners of an equilateral triangle of side length $d_\mr{BS}$, similar to \cite{brandt2014_hardwareimpaired_wmmse} without directional antennas.
	For each channel realization, all \glspl{ue} are randomly and uniformly placed inside the sextant centered on their assigned \gls{bs} with radius $d_\mr{BS}/\sqrt{3}$.
	The large-scale path loss $PL_{ik}$ between \gls{bs} $k$ and \gls{ue} $i$ is calculated according to the picocell model \cite{lopezperez2012_3gpp_mobility_management}.
	We assume a rich scattering environment for both the \glspl{bs} and \glspl{ue} and Rayleigh fading.
	Thus, the channel matrix coefficients $\matel{\matr{H}_{ik}}{nm}$ are sampled from $\cnormdistr{0, 10^{\frac{PL_{ik}}{\SI{10}{\decibel}}}}$.
	In Section \ref{sec:main:deepmimo} and Section \ref{sec:main:finetuning}, the DeepMIMO dataset \cite{alkhateeb2019_deepmimo} is leveraged instead.
	For simplicity, we assume equal \gls{ue} antennas dimensions $N_i = N$, \gls{ue} noise variances $\sigma_i^2 = \sigma_\mr{UE}^2$, sum-rate weights $\alpha_i=1$, \gls{bs} antenna dimensions \gls{bs} $M_k = M$ and \gls{bs} power budget $P_k= P_\mr{BS}$. 
	Similarly, an equal amount of \glspl{ue} $\abs{\setik} = \mathcal{I}$ is assigned to each individual \gls{bs}, however, GCN-WMMSE exhibits comparable performance if $\abs{\setik}$ is not equal for all $k$.
	\par
	\begin{table}[t]
		\centering
			\centering
			\caption{Base parameter set for GCN-WMMSE networks and general training parameters. \label{tab:main:exp_base_modelparam}} 
			\begin{tabularx}{\linewidth}{Xr}
				\toprule
				\multicolumn{2}{l}{\textbf{Base Model Hyperparameters}}\\
				\midrule
				Number of Layers $L$ & $7$ \\
				Polynomial Degree $G$ & $2$ \\
				Number of Filters $F$ & $4$ \\
				\midrule
				\multicolumn{2}{l}{\textbf{Base Training Hyperparameters}}\\
				\midrule
				Loss Function & \eqref{eq:main:normloss} with $\mathcal{L}=\matset{L}$ \\
				ADAMW $(\beta_1, \beta_2, \lambda)$ & $(0.9, 0.99, 10^{-3})$\\
				Learning Steps $T$ & $10^4$\\
				Learning Rate $\eta$ & $0.01$, $/10$ after every 2500 steps\\
				Minibatch Size $\abs{\mathcal{T}_t}$ & 100\\
				Gradient Clipping Value & $1$ \\
				\bottomrule
			\end{tabularx}
	\end{table}
	\begin{table}
			\centering
			\caption{Base scenario configuration for the validation and training scenario samples.\label{tab:main:exp_base_scenarioparam}}
			\begin{tabularx}{\linewidth}{Xr}
				\toprule
				\multicolumn{2}{l}{\textbf{Base Scenario Configuration}}\\
				\midrule
				BS Distance $d_\mr{BS}$ & \SI{200}{\meter} \\
				BS Antenna Dimension $M$ & $12$ \\
				BS Tx Power $P_\mr{BS}$ & \SI{30}{\dBm}\\
				Num. of UEs $I$ & $12$, equal num. assigned per \gls{bs} \\
				UE Antenna Dimension $N$ & $2$ \\
				UE Noise Power $\sigma^2_\mr{UE}$ & \SI{-100}{\dBm}\\
				\bottomrule
			\end{tabularx}
	\end{table}
	All experiments from Section \ref{sec:main:generalperf} to \ref{sec:main:finetuning} are conducted using the network and training hyperparameters summarized in \tab{tab:main:exp_base_modelparam}, unless specified otherwise.
	The filter taps of the downlink beamformer \glspl{gcn} are initialized according to \cite{trabelsi2017_deep_complex_networks}, the biases with $\nullvec$ and the taps of the weight \glspl{gcf} by $1/(G+1)$.
	GCN-WMMSE is initialized by normalized \gls{mrc} beamformers $\matr{V}_i\itindex{0} \propto \matr{H}_{ik}\herm$ for $i \in \mathcal{I}_k$.
	The Lagrangian variable is initialized as $\mu_k\itindex{\ell,0}=10^{-12}$ for every $\ell$ and updated for ${P}=8$ iterations.
	The validation sets contain $1000$ scenario realizations but the training samples are 'single-use' and newly sampled at each step $t$, unless specified otherwise.
	We compare to the classical \gls{wmmse} algorithm. 
	WMMSE RI denotes the \gls{wsr} achieved by the \gls{wmmse} algorithm averaged over $50$ random initializations per scenario realization. 
	Random beamformer initializations are obtained by sampling the matrix elements from $\cnormdistr{0,1}$ and normalization such that the power constraints are exactly met.
	We earmark WMMSE50 as the best \gls{wsr} achieved by the \gls{wmmse} algorithm over these $50$ initializations.
	It serves as a benchmark, however, we remark that it is impractical due to its high computational cost.
	\gls{wmmse} \gls{mrc} denotes the \gls{wmmse} algorithm initialized by \gls{mrc} beamformers.
	The \gls{wmmse} algorithm is always carried out for $100$ iterations. 
	The acronym TR indicates the average rate that the \gls{wmmse} algorithm achieves when being truncated after $L$ iterations, therefore, requiring the same communication overhead as GCN-WMMSE.
	The average achievable \gls{wsr} is denoted by $\rate_\Sigma$, and $\rate_\Sigma^\mr{rel.}$ denotes the average \gls{wsr} relative to the result of WMMSE50 in percent.
	Tab. \ref{tab:main:overview_algs} provides an overview over the variants of the algorithm and the proposed model.
	\begin{table}
		\caption{Overview of compared algorithm and model variants.\label{tab:main:overview_algs}}
		\begin{tabularx}{\linewidth}{lX}
			\toprule
			\textbf{Acronym} & \textbf{Explanation}\\
			\midrule
			\textbf{GCN-WMMSE} & Proposed architecture...\\
			~~~~~MT & ...trained on scenarios configured as in test set.\\
			~~~~~PT & ...trained on scenario configuration marked by arrow.\\
			~~~~~DT & ...trained on random scenario configuration.\\
			\midrule
			\textbf{WMMSE} & Baseline algorithm...\\
			~~~~~RI & ...with random initializations.\\
			~~~~~RI TR & ...with random initializations truncated to $L$ iterations.\\
			~~~~~50 & ...with best result of 50 initializations\\
			~~~~~MRC & ...with \gls{mrc} initialization.\\
			~~~~~MRC TR & ...with \gls{mrc} initialization truncated to $L$ iterations.\\
			\bottomrule
		\end{tabularx}
	\end{table}
	\subsection{General Performance and Ablation Study} \label{sec:main:generalperf}
	\par
	In this section, GCN-WMMSE networks are trained and validated on scenarios configured as in \tab{tab:main:exp_base_scenarioparam}.
	\begin{table}[t]
			\centering
			\caption{Absolute and relative rate of the proposed GCN-WMMSE networks compared to WMMSE50 over network depths $L$. \label{tab:main:exp_sweep_num_layers}}
			\begin{tabularx}{\linewidth}{X @{\hspace{1\tabcolsep}} r @{\hspace{1\tabcolsep}} r @{\hspace{1\tabcolsep}} r @{\hspace{1\tabcolsep}} r @{\hspace{1\tabcolsep}}r @{\hspace{1\tabcolsep}} r @{\hspace{1\tabcolsep}} r} %
				\toprule
				\textbf{Number of Layers} $L$ & 3 & 4 & 5 & 6 & 7 & 8 & 9\\
				\midrule
				Abs. Rate $\rate_\Sigma$ (\si{\nats}) & 82.21 & 87.84 & 89.98 & 91.67 & 92.56 & 93.46 & 93.89\\
				Rel. Rate $\rate_\Sigma^\mr{rel.}$ (\%) & 83.72 & 89.45 & 91.63 & 93.35 & 94.26 & 95.17 & 95.61\\
				\bottomrule
			\end{tabularx}
	\end{table}
	\subsubsection{Number of Network Layers $L$}
	\tab{tab:main:exp_sweep_num_layers} shows that the rate of the proposed GCN-WMMSE increases consistently from $3$ to $9$ network layers up to $95.61\%$ relative \gls{wsr}, however, there are significant diminishing returns after about $6$ layers.
	Nevertheless, since \gls{wmmse} RI achieves $95.20\%$ relative \gls{wsr} only after 100 iterations, the results prove that the computation time and iterations can be drastically reduced.
	We select $L=7$ for the following experiments.
	\subsubsection{Number of GCN Filter Features $F$}
\begin{figure}[t]
		\setlength{\abovecaptionskip}{-12pt}%
		\input{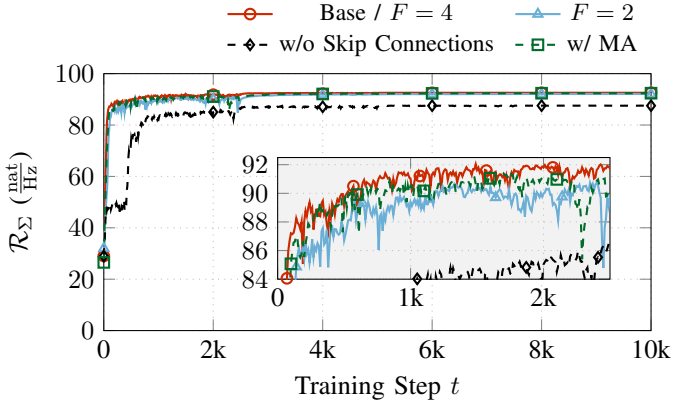}
		\caption{Convergence of GCN-WMMSE models evaluated on validation set over $T$ training steps.\label{fig:main:training_over_steps}}%
	\end{figure}

	\begin{table}[t]
			\centering
			\caption{Rate and relative rate of the proposed GCN-WMMSE networks compared to WMMSE50 for different numbers of filters $F$. \label{tab:main:exp_sweep_vchannel}}
			\begin{tabularx}{\linewidth}{Xrrrr}
				\toprule
				\textbf{Number of Filters} $F$ & 1 & 2 & 4 & 6\\
				\midrule
				Abs. Rate $\rate_\Sigma$ (\si{\nats}) & 91.49 & 92.08 & 92.56 & 92.56\\
				Rel. Rate $\rate_\Sigma^\mr{rel.}$ (\%) & 93.16 & 93.77 & 94.26 & 94.26\\
				\bottomrule
			\end{tabularx}
	\end{table}
	\begin{table}
		\centering
		\caption{Ablation study of the architecture components of GCN-WMMSE (proposed). \label{tab:main:ablation}}
		\begin{tabularx}{\linewidth}{Xrr}
			\toprule
			\textbf{Changed Component} & $\rate_\Sigma$ (\si{\nats}) & $\rate_\Sigma / \rate_{\Sigma,\mr{Base}}$ (\%)\\
			\midrule
			Base & 92.56 & 100.00\\
			\midrule
			w/o Weight Matrix GCF & 91.72 & 99.09 \\
			w/o Skip Connections & 87.53 & 94.57 \\
			w/o Biasing/$\modrelu$ & 89.20 & 96.37 \\
			w/o Diagonal Loading & 92.70 & 100.15 \\
			w/ MA & 92.39 & 99.81 \\
		\end{tabularx}
	\end{table}
	\tab{tab:main:exp_sweep_vchannel} shows that increasing the number of \gls{gcn} filters $F$ in each layer only yields a marginal increase in the \gls{wsr}.
	An increase of $F$ is, however, associated with an increase in the number of `good' filter initializations, which leads to accelerated and stabilized training as \fig{fig:main:training_over_steps} demonstrates.
	
	\begin{figure}
		\setlength{\abovecaptionskip}{-12pt}%
			\input{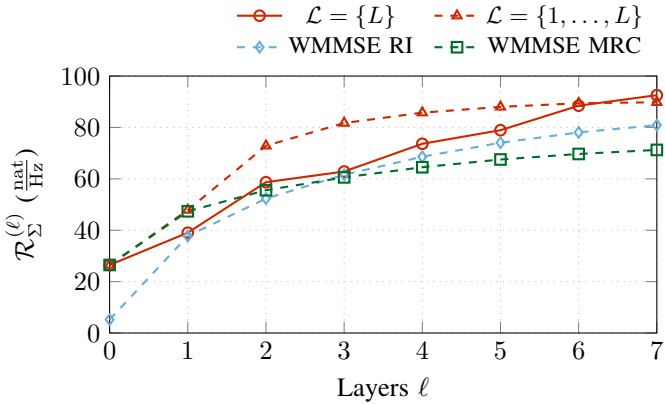}
			\caption{Performance of GCN-WMMSE (proposed) given the layer outputs $\matr{V}_i\itindex{\ell}$ for a network with $L=7$ layers. The loss \eqref{eq:main:normloss} is based on the output of either the last or all layers.}%
		\label{fig:perfeval:exp_layers_all_last}
	\end{figure}
	\subsubsection{Ablation Study}
	In \tab{tab:main:ablation}, individual network components are added or removed one at a time to study their impact on the \gls{wsr} at the last layer.
	In case of the weight matrix \gls{gcf} (\cf \eqref{eq:main:weightgcf}), it is replaced by the original \gls{wmmse} update \eqref{eq:theory:wupdate}.
	\textit{Diagonal Loading} and \textit{Biasing} refers to the terms with the parameters $\vect{a}_{\mr{V},\ell 0}$ and $\vect{b}_\ell$ in \eqref{eq:main:gcnupdate}.
	MA (moving average) refers to an additional term in \eqref{eq:main:gcnupdate} containing the non-inverted loaded weighted uplink covariance matrix $\tilde{\matr{R}}_k\itindex{\ell}$, \ie, we have
	$
		\tilde{\matr{P}}_{id}\itindex{\ell} = (\tilde{\matr{R}}_k\itindex{\ell})^\dagger \vvtilde_{id}\itindex{\ell} \vect{a}_{\mr{V},\ell 1}\tran + \vvtilde_{id}\itindex{\ell}\vect{a}_{\mr{V},\ell 0}\tran + \tilde{\matr{R}}_k\itindex{\ell} \vvtilde_{id}\itindex{\ell} \vect{a}_{\mr{ma},\ell}\tran
	$
	with an additional trainable parameter $\vect{a}_{\mr{ma},\ell}\in \Cset^F$.
	\par
	The results are given in \tab{tab:main:ablation} and have a two-sided $99\%$-confidence of $\pm\SI{0.1}{\nats}$.
	The biasing, which activates the $\modrelu$-nonlinearity, and the skip connections have the largest performance impact.
	The effectiveness of skip connections is due to a significant acceleration and stabilization in training \cite{he2015_resnet, srivastava15_highwaynetworks} as shown in \fig{fig:main:training_over_steps}.
	The weight matrix \gls{gcf} has a statistically significant though comparably low impact.
	Its computational cost, however, is low and it can be empirically shown that it becomes more advantageous for higher \gls{snr}.
	On the other hand, the results suggest that the diagonal loading term could be detrimental. %
	Adding an MA-term does not improve the performance in this setup.
	
	\subsubsection{Loss Layer Set $\mathcal{L}$}
	\fig{fig:perfeval:exp_layers_all_last} investigates the impact of the choice of layers $\ell$ in the layer set $\mathcal{L}$ considered in the loss function in \eqref{eq:main:normloss}.
	When performing greedy \gls{sgd} on all layers, a rate of \SI{81.71}{\nats} ($83.21\%$ compared to WMMSE50) is achieved after only $3$ layers, \gls{wmmse} RI requires $8$ iterations to achieve a similar result.
	However, the network performance saturates as the greedy method risks slipping into inferior local optima, while $\mathcal{L} = \matset{L}$ eventually achieves a higher rate of \SI{92.56}{\nats} (compared to \SI{89.81}{\nats}).
	In a deployment scenario with a flexible runtime or number of feedback loops, \ie exchanges between transmitters and receivers, the algorithm may be terminated early after a few layers.
	Therefore, the trade-off in choosing $\mathcal{L}$ is important to consider.
	
	\subsubsection{Training Data Set Size $\abs{\mathcal{T}}$}
	\begin{table*}
		\centering
		\caption{Absolute and relative \gls{wsr} of GCN-WMMSE (proposed) for a finite set of training samples. \label{tab:main:training_set_size}}
		\begin{tabularx}{0.8\linewidth}{Xrrrrrrrrrr}
			\toprule
			Training Set Size $\abs{\mathcal{T}}$ & 100 & 200 & 300 & 400 & 500 & 600 & 700 & 800 & 900 & 1000\\
			\midrule
			Abs. Rate $\rate_\Sigma$ (\si{\nats}) & 91.58 & 91.82 & 91.73 & 92.06 & 92.17 & 92.08 & 92.09 & 92.28 & 92.50 & 92.16\\
			Rel. Rate $\rate_\Sigma^\mr{rel.}$ (\%) & 93.26 & 93.50 & 93.41 & 93.75 & 93.85 & 93.77 & 93.78 & 93.97 & 94.19 & 93.84\\
			\bottomrule
		\end{tabularx}
	\end{table*}
	Finally, instead of training data being generated on demand, a finite training data set is considered in \tab{tab:main:training_set_size}.
	We observe that for a data set size of $100$ scenario samples the relative \gls{wsr} already achieves a value of $93.26\%$. %
	This remarkable data efficiency is enabled by the equivariance properties of the graph filter structures and the preservation of the original \gls{wmmse} algorithm structure. 
	The number of parameters is low with only $229$ scalar network parameters in this example.
	\subsubsection{Computational Cost and Communication Overhead}
		\begin{table}
		\centering
		\caption{Comparison of communication overhead and computational cost between GCN-WMMSE (proposed) and WMMSE algorithm.\label{tab:main:computational_cost}}
		\newcolumntype{Y}{>{\raggedleft\arraybackslash}X}
		\begin{tabularx}{\linewidth}{YYYY}
			\toprule
			\multicolumn{2}{c}{\textbf{\# Communication Rounds}} & \multicolumn{2}{c}{\textbf{Computational Cost} [ms]}\\\cmidrule(lr){1-2}\cmidrule(lr){3-4}
			GCN-WMMSE & WMMSE RI & GCN-WMMSE & WMMSE RI\\
			\midrule
			6 & 42 & 3.07 & 17.82 \\
			7 & 63 & 3.57 & 26.70\\
			8 & 98 & 4.12 & 41.83 \\
			\bottomrule
		\end{tabularx}
	\end{table}
	We compare the computational cost and communication overhead of GCN-WMMSE for $L=6,7,8$ against WMMSE RI with a number of iterations that achieves a comparable \gls{wsr} on the base scenario.
	Note that the communication overhead is proportional to the number of required communication rounds, which itself is equal to the number of layers/iterations.
	The computational cost is given for a centralized execution with a AMD R7 2700X CPU with 32GB memory.
	\par
	\tab{tab:main:computational_cost} demonstrates that GCN-WMMSE reduces the computational cost by a factor of $6$ up to $10$, with the relative advantage increasing for deeper networks.
	Simultaneously, the number of required communication rounds decreases by roughly an order of magnitude.
	The per-iteration computational cost increase is $\sim 20\%$ in our modular implementation.
	\subsection{Generalization Capabilities}  \label{sec:main:generalization}
	This subsection investigates the generalization capabilities of the proposed GCN-WMMSE architecture.
	The scenario configuration for the training set and the validation set is given in \tab{tab:main:exp_base_scenarioparam} unless defined otherwise.
	We study the relative achievable \gls{wsr} of GCN-WMMSE networks by varying individual scenario parameters (i) given a network trained on samples with matching scenario configuration to the validation data, denoted by MT, (ii) given a network which is only trained on samples at a defined pivot scenario configuration, denoted by PT, and (iii) given a network which uses a training set containing random scenario configurations, denoted by DT.
	The achievable \gls{wsr} of the WMMSE algorithm with \gls{mrc} initialization is not provided in this section since it is consistently outperformed by WMMSE RI in our experiments. 
	\begin{figure*}[t]
		\centering
		\input{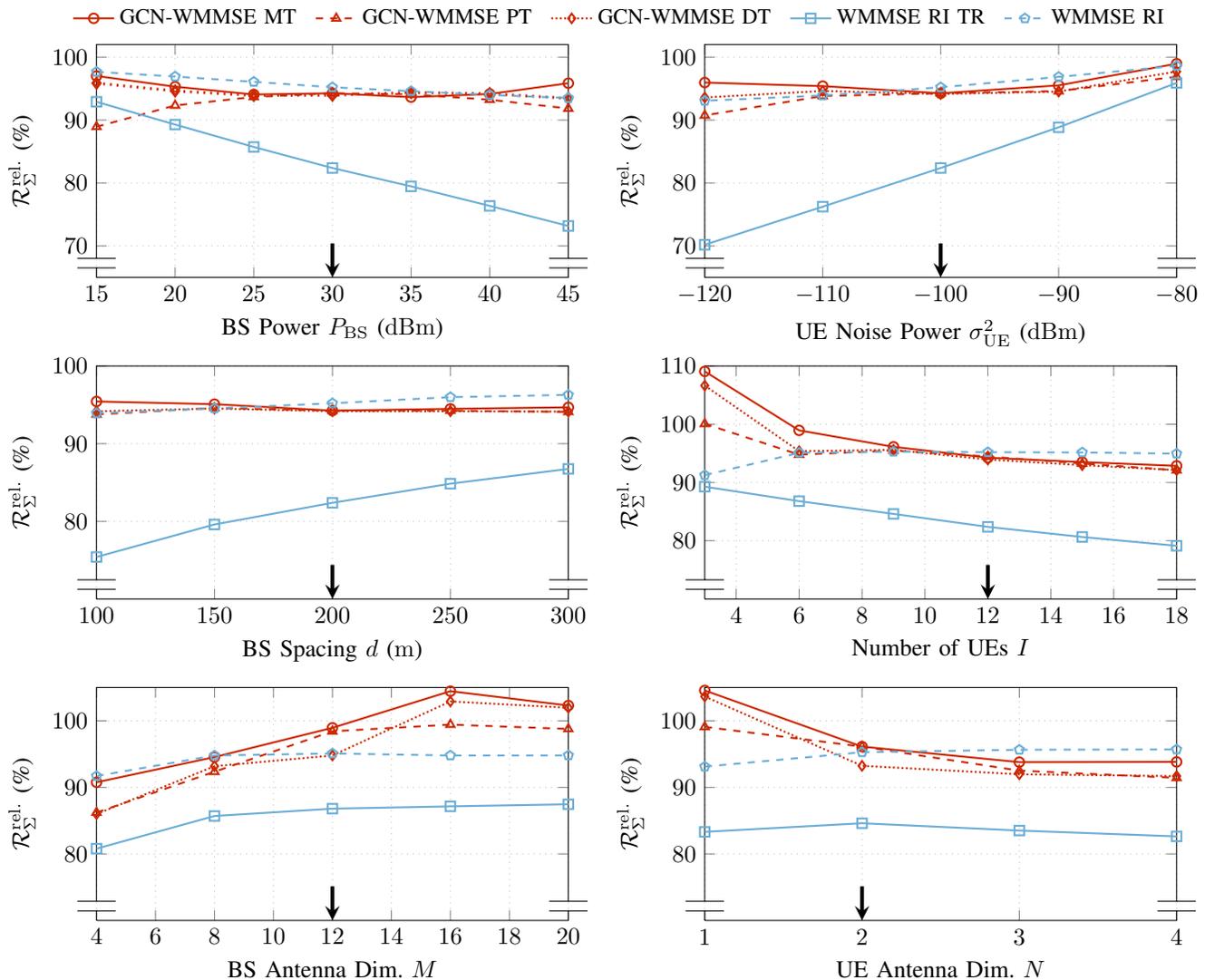}
		\caption{Generalization of GCN-WMMSE (proposed) \wrt several wireless scenario parameters. The arrows indicate the configuration for the training of GCN-WMMSE PT networks. Top: \gls{bs} power budgets $P_k$ (left) and \gls{ue} noise power $\sigma_\mr{UE}^2$  (right). Middle: \gls{bs} separation distances $d_\mr{BS}$ (left) and number of \glspl{ue} (right). Bottom: Numbers of antenna elements at the \glspl{bs} (left) and at the \glspl{ue} (right).}
		\label{fig:main:exp_mimo_generalization}
		\vspace{-10pt}%
	\end{figure*}%
	
	\subsubsection{BS Power and UE Noise Power}
	In \fig{fig:main:exp_mimo_generalization}, we sweep the \gls{bs} power budget (top left) or \gls{ue} noise power (top right), while GCN-WMMSE PT is trained with $P_\mr{BS} = \SI{30}{\dBm}$ or $\sigma_\mr{UE}^2 = \SI{-100}{\dBm}$, respectively.
	GCN-WMMSE MT consistently achieves over $93.64\%$ relative \gls{wsr}, closely following WMMSE RI and outperforming the WMMSE RI TR significantly, especially for high values of $P_\mr{BS}$.
	GCN-WMMSE PT generalizes well to scenarios with higher \gls{bs} power compared to scenarios in its training data, however, it sacrifices some relative \gls{wsr} for lower \gls{bs} power.
	In case of a $\SI{-15}{\decibel}$ offset from the training data, the relative \gls{wsr} achieved by GCN-WMMSE PT falls below the value of WMMSE RI TR.
	On the other hand, the achievable \gls{wsr} for rising \gls{snr} remains close to WMMSE RI.
	For varying noise levels $\sigma_\mr{UE}^2$, both GCN-WMMSE MT and GCN-WMMSE PT consistently outperform WMMSE RI TR, and GCN-WMMSE MT even outperforms WMMSE RI in case of low noise. 
	
	\subsubsection{Network Density}
	\fig{fig:main:exp_mimo_generalization} (middle left) illustrates the generalization \wrt the \gls{bs} distance $d_\mr{BS}$. 
	GCN-WMMSE PT is trained with $d_\mr{BS}=\SI{200}{\meter}$.
	In this case, both the GCN-WMMSE MT and PT networks closely follow the \gls{wsr} of WMMSE RI, significantly above the truncated algorithm and achieving rates above $93.76\%$ relative to WMMSE50.
	This can be explained by the \gls{sinr} at the \glspl{ue} staying approximately constant if $d_\mr{BS}$ changes and the interference is significant.
	\subsubsection{Array Dimensions}
	In \fig{fig:main:exp_mimo_generalization} (bottom left), we study a changing \gls{bs} antenna dimension with $I=6$ \glspl{ue}.
	GCN-WMMSE PT is trained with $M=12$.
	GCN-WMMSE MT performs well for $M<12$ and substantially outperforms WMMSE RI with 100 iterations for $M\geq12$.
	In this case, the classical WMMSE algorithm tends to find suboptimal beamformers with significant differences between individual \gls{ue} rates per scenario realization while the unfolded algorithm favors solutions with uniformly distributed rates.
	GCN-WMMSE PT outperforms WMMSE RI for $M\geq12$ as well, but to a lesser extent.
	\par
	Varying numbers of \gls{ue} antenna elements are considered in \fig{fig:main:exp_mimo_generalization} (bottom right) with $I=9$ \glspl{ue}.
	GCN-WMMSE PT and GCN-WMMSE MT outperform WMMSE RI TR, both achieving a rate above $91.43\%$ relative to WMMSE50.
	For $N=1$, WMMSE RI is substantially outperformed.
	\par
	These generalization capabilities can be leveraged in training to facilitate training data collection.
	Instead of sharing the full channel matrices, truncated channels can be collected at a learning node to reduce communication overhead and to decrease the computational cost in online training applications, see Section \ref{sec:main:finetuning}.
	\par
	We remark that both the WMMSE and GCN-WMMSE can experience numerical difficulties when the number of \gls{bs} antennas $M$ exceeds the total number of antennas of all \glspl{ue}.
	In this case, if the candidate beamformers $\vmtilde_i$ are additionally (almost) orthogonal to the nullspace of $\rmat_k$ and it holds that $\sum_{i \in \setik} ||\rmat_k^\dagger \vmtilde_i||_F^2 \leq P_k$, the minimization of \eqref{eq:main:reform_sum_rate_problem} \wrt $\vm$, a \gls{qcqp} which is solved by the updates \eqref{eq:theory:dualfun} and \eqref{eq:theory:vupdate}, becomes ill-conditioned.
	In literature, this is known as the \textit{(near) hard case} of a \gls{qcqp} \cite{rojas2001_trustregion}.
	Our results indicate that GCN-WMMSE is more robust than the WMMSE algorithm in wireless scenarios prone to this issue.
	Additional remedies include replacing the operations \eqref{eq:theory:dualfun} and \eqref{eq:theory:vupdate} with a specialized iterative solver that achieves a higher accuracy in such instances \cite{rojas2001_trustregion}.
	Note that ill-conditioned instances do not occur if $K=1$.
	
	\subsubsection{Number of UEs}
	Lastly, \fig{fig:main:exp_mimo_generalization} (middle right) shows the achieved relative \gls{wsr} given a varying number of \glspl{ue} $I$.
	GCN-WMMSE MT generalizes well, approximately matching or outperforming WMMSE RI on average. %
	GCN-WMMSE PT is trained on data with 12 \glspl{ue} and generalizes well to lower $I$, achieving a $\rate_\Sigma^\mr{rel.}$ of $96.09\%$. %
	For $I=18$ \glspl{ue}, it outperforms WMMSE RI TR by $13\%$.
	It is thus advantageous to train with the maximum number of \glspl{ue} disregarding the increased complexity of training.
	\subsubsection{Training on Random Scenario Parameters}
	Instead of relying on pure transfer learning, \ie, if training and test data have different statistics, GCN-WMMSE DT leverages datasets containing random scenario parameters.
	As \fig{fig:main:exp_mimo_generalization} demonstrates, this is advantageous in case of the \gls{bs} power or \gls{ue} receiver noise, and similar in performance in case of diverse \gls{bs} distances. 
	For changing antenna dimensions, this training scheme is beneficial compared to GCN-WMMSE PT when the number of \gls{bs} antennas is high or in the case of single-antenna receivers.
	Similarly, it outperforms pure transfer learning for a low number of \glspl{ue}.
	\subsection{DeepMIMO Dataset} \label{sec:main:deepmimo}
	The DeepMIMO dataset \cite{alkhateeb2019_deepmimo} contains precomputed, ray-traced \gls{csi} for an urban scenario.
	It offers multiple \glspl{bs} and thousands of possible \glspl{ue} positions.
	The dataset enables the evaluation of the proposed GCN-WMMSE architecture for correlated channel coefficients.
	\par
	\begin{table*}[t]
	\centering
	\caption{Deep \gls{mimo} scenario configuration using dataset `O1'.}
	\label{tab:main:deepmimo_param}
	\begin{tabularx}{14cm}{Xrrrrr}
		\toprule
		\textbf{Scenario Configuration} & DM S1 & DM S2 & DM S3 & DM S4a & DM S4b\\
		\midrule
		Active \glspl{bs} & 13, 14, 15, 16 & 5, 6, 7, 8 & 6, 8, 17, 18 & 13, 14, 15, 16 & 13, 14, 15, 16  \\
		Active \gls{ue} Rows & 2752 to 3600 & 1400 to 1900 & 3853 to 4750 & 2752 to 3600 & 3300 to 3400\\
		Number of active \glspl{ue} & 16 & 16 & 16 & 4 & 12\\
		Number of \gls{bs} Antennas & $4\times 3 \times 1$ & $4\times 3 \times 1$  & $4\times 3 \times 1$ & $4\times 3 \times 1$ & $4\times 3 \times 1$\\
		Number of \gls{ue} Antennas & 1 & 1 & 2 & 4 & 4\\
		\bottomrule	
	\end{tabularx}
	\end{table*}
	Three specific scenario configurations, denoted as DM S1, DM S2 and DM S3 respectively, with dataset parameters as in \tab{tab:main:deepmimo_param} are defined. 
	Furthermore, we set $P_\mr{BS}=\SI{30}{\dBm}$, $\sigma_\mr{UE}^2=\SI{-90}{\dBm}$, $B=\SI{240}{\kilo \hertz}$, and consider a single OFDM carrier.
	The \gls{bs} antennas are spaced by half a wavelength.
	For each configuration the set of \gls{ue} positions is split in two halves which are then assigned to their closest \glspl{bs}. 
	Afterwards, $5000$ random sets of $16$ \glspl{ue}, $3$ per \gls{bs}, are sampled as training data, and $1000$ sets are sampled as validation data.
	The sets of \gls{ue} positions in training and validation data are disjoint.
	
	\par
	\begin{table*}[t]
	\centering
	\caption{Rate and relative rate of GCN-WMMSE (proposed) compared to WMMSE50 on the Deep MIMO data set scenarios.} %
	\label{tab:main:deepmimo_results}
	\begin{tabularx}{15cm}{Xrrrr}
		\toprule
		\multirow{2}{*}{\parbox{2cm}{\textbf{Scenario} \\ \textbf{Configuration} }} & \multicolumn{3}{r}{Rel. Rate $\rate_\Sigma^\mr{rel.}$ (\%)} & Abs. Rate $\rate_\Sigma$ (\si{\nats})\\
		\cmidrule(lr){2-4} \cmidrule(lr){5-5}
		& GCN-WMMSE & WMMSE RI & WMMSE RI TR & WMMSE 50\\
		\midrule
		DM S1 & 93.47 & 91.76 & 78.68 & 38.61 \\
		DM S2 & 91.60 & 90.63 & 75.62 & 49.63 \\
		DM S2 (trained on DM S1) & 90.05 & 90.63 & 75.62 & 49.63\\
		DM S3 & 89.43 & 93.43 & 78.45 & 50.06 \\
		\midrule
		DM S4b & 90.23 & |~~~ & |~~~ & |~~~  \\
		DM S4b (trained on DM S4a) & 86.35 & 94.21 & 81.88 & 35.61  \\
		DM S4b (trained on DM S4a, finetuned) & 88.29 & |~~~ & |~~~ & |~~~ \\
		\bottomrule
	\end{tabularx}
	\end{table*}
	We train the GCN-WMMSE networks similarly to the previous experiments with hyperparameters summarized in \tab{tab:main:exp_base_modelparam} while the number of layers $L$ is set to $5$.
	As \tab{tab:main:deepmimo_results} shows, the proposed GCN-WMMSE networks achieve at least $89.43\%$ of WMMSE50, outperforming the WMMSE RI TR with $L=5$ iterations by at least $10\%$.
	It improves over WMMSE RI in DM S1 and DM S2, but falls $4\%$ short in DM S3.
	In the latter scenario, increasing the number of layers improves the performance.
	Furthermore, a network model trained on DM S1 generalizes with little performance loss to DM S2.
	Overall, the benefit of a significantly reduced number of required iterations observed in the Rayleigh fading scenario transfers to the ray-tracing channel model.
	\subsection{Case Study: Finetuning in Dynamic Scenarios} \label{sec:main:finetuning}
	Although GCN-WMMSE exhibits significant generality, finetuning of a network model by learning on current data may still be beneficial in case of changing wireless conditions.
	Therefore, it is intriguing to leverage the characteristics of GCN-WMMSE to enable efficient finetuning.
	To verify this idea, we consider two DeepMIMO scenarios S4a and S4b, see \tab{tab:main:deepmimo_param}.
	Contrary to S4a, scenario S4b features a number of \glspl{ue} tightly clustered in a hotspot.
	A GCN-WMMSE model is then trained on S4a, after which the model is transferred to scenario S4b and finetuned for $T=250$ steps.
	Given local CSI availability, it is of interest to reduce the transfer of channel samples to a training node to a minimum.
	We thus limit the number of samples available to the finetuning stage to $100$ to take advantage of the training data efficiency of GCN-WMMSE.
	Furthermore, we truncate channel matrices from $4\times 12$ to $3\times9$ to leverage generality in the antenna dimension, almost halving the size of the channel data.
	This additionally reduces the computational cost of training.
	All other simulation parameters are as in Section \ref{sec:main:deepmimo}.
	\par
	\tab{tab:main:deepmimo_results} demonstrates that finetuning with the small truncated dataset can improve the performance by almost $2\%$ over pure model transfer, which approaches the performance of the model trained on S4b.
	Thus, the properties of GCN-WMMSE enable efficient online training with distributed data collection.
	
	\section{Related Work}\label{sec:relatedwork}
	In this section, we discuss previous model architectures based on WMMSE algorithm unrolling.
	We focus on the works \cite{chowdhury2021_unfolding,pellaco2020_deep_unfolding_pgd,hu2021_iadnn} specifically.
	\tab{tab:main:overview_relwork} provides an overview.
	These works limit the network domain compared to the classical \gls{wmmse}, and particularly the architectures considered in \cite{chowdhury2021_unfolding,hu2021_iadnn} cannot easily be extended to the general setup considered in this paper.
	We thus compare to the proposed GCN-WMMSE on the respective limited scenario setup.
	If not mentioned otherwise, all networks are trained using the training hyperparameters in \tab{tab:main:exp_base_modelparam}.
	\begin{table}
		\caption{Overview of a selection of related architectures.\label{tab:main:overview_relwork}}
		\begin{tabularx}{\linewidth}{lXX}
			\toprule
			\textbf{Acronym} & \textbf{Explanation} & \textbf{Domain}\\
			\midrule
			UWMMSE \cite{chowdhury2021_unfolding} & Integration of \gls{gcn}. & SISO Tx-Rx-pairs.\\
			PGD WMMSE \cite{pellaco2020_deep_unfolding_pgd} & \gls{pgd} with learnable step size. & Extendable to cellular MU-MIMO.\\
			IAIDNN \cite{hu2021_iadnn} & Trainable 1st-order approximation. & Single-cell MU-MIMO.\\
			\bottomrule
		\end{tabularx}
	\end{table}
	\subsection{UWMMSE}
	The authors of \cite{chowdhury2021_unfolding} introduce the UWMMSE (Unfolded WMMSE) architecture which is limited to transmitter-receiver pairs ($K=I$) and optimization of power allocation on \gls{siso} links ($M=N=1$).
	The problem reduces to real variables and problem \eqref{eq:theory:dualfun} significantly simplifies.
	The architecture aims to reduce the number of required iterations compared to the original \gls{wmmse} algorithm by transforming the weight scalars $w_i$ (1-dimensional weight matrix $\wm_i$) by an affine mapping.
	The mapping coefficients are obtained by 2-layer \glspl{gcn} with a shift matrix consisting of all channels and trainable input vectors.
	For more details, see \cite{chowdhury2021_unfolding}.
	\par
	Models based on both the UWMMSE and proposed GCN-WMMSE architecture are compared on \gls{siso} scenarios with a similar channel model as in \cite[Section 4E]{chowdhury2021_unfolding}:
	For each channel sample, $K$ transmitters are placed randomly at a location $\vect{p}_{\mr{Tx},i} \in \left[-K/d, K/d\right]^2$ and their paired receiver is placed at random at $\vect{p}_{\mr{Tx},i} + \vect{o}_{\mr{Rx},i}$ where $\vect{o}_{\mr{Rx},i} \in\left[ -M/4, M/4\right]^2$. 
	Parameter $d$ denotes the network density. 
	The scalar channel is given by $h_{ij} = \abs{\vect{o}_{\mr{Rx},i}}^{-2.2} h_{\mr{f}, ij}$ where $h_{\mr{f}, ij}$ is drawn from a Rayleigh distribution with mode $1$.
	The large-scale path loss $\abs{\vect{o}_{\mr{Rx},i}}^{-2.2}$ is bounded by $1$ from above. %
	\begin{figure*}[t]
		\input{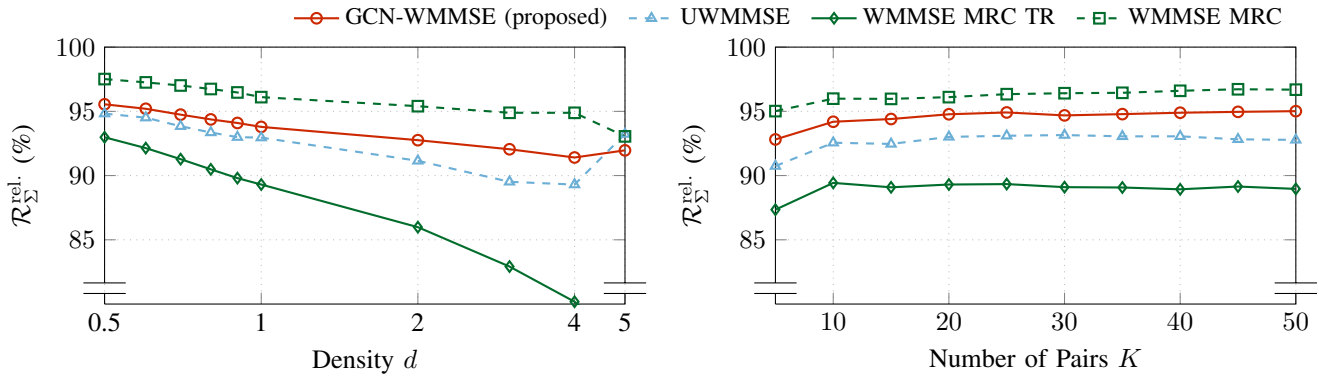}
		\caption{UWMMSE \cite{chowdhury2021_unfolding} and GCN-WMMSE (proposed) for varying density $d$ (left) or number of pairs $K$ (right). $P_\mr{BS} = \SI{0}{\dBm}$, $\sigma_\mr{UE}^2 = \SI{-92}{\dBm}$.} 
		\label{fig:relatedwork:uwmmse}
		\vspace{-10pt}
	\end{figure*}
	\par
	For the following experiments, the number of layers $L$ of both networks is set to $4$.
	The UWMMSE network uses $2$-layered \glspl{gcn} with a hidden layer dimension of $4$, the GCN-WMMSE models adopt $F$ and $G$ from \tab{tab:main:exp_base_modelparam}.
	First, following \cite{chowdhury2021_unfolding}, both networks are trained on randomly sampled channels with fixed $M=20$ transceiver pairs and density drawn from $[0.5, 5]$ uniformly at random.
	The networks are validated on a sample set with fixed density $d$ and $1000$ samples each.
	\fig{fig:relatedwork:uwmmse} (left) shows that both networks outperform the truncated \gls{wmmse} with \gls{mrc} initalization, with the gap increasing for denser and more difficult scenarios.
	We only show \gls{wmmse} \gls{mrc} since it consistently outperforms \gls{wmmse} RI in these scenarios.
	GCN-WMMSE outperforms the \gls{siso}-specific UWMMSE with an exception at density $d=5$, where UWMMSE closely approaches the converged \gls{wmmse}. %
	Note that for $d=1$, UWMMSE and GCN-WMMSE achieve a rate of $\SI{50.14}{\nats}$ and $\SI{50.59}{\nats}$, respectively; the classical \gls{wmmse} algorithm only approaches a similar result after $11$ or $17$ iterations, respectively.
	\par
	We repeat the experiment with fixed $d = 1$ and instead vary the number of pairs $K$ from $5$ to $50$.
	As indicated in \fig{fig:relatedwork:uwmmse} (right),
	GCN-WMMSE outperforms UWMMSE, achieving consistently between $92\%$ to $95\%$ of WMMSE50, ahead of WMMSE MRC TR.
	In conclusion, the proposed GCN-WMMSE architecture generalizes well to pairwise \gls{siso} scenarios, and outperforms the less general scenario-configuration-specific architecture in most cases.
	Our results for UWMMSE networks match the results reported in the original work \cite{chowdhury2021_unfolding}. Recently, the same authors published an extension of UWMMSE for \gls{mimo} links \cite{chowdhury2021_mlaided_tacticalmimo}, however, compared to GCN-WMMSE it is restricted to pairwise links and real channels and parameter sets are unflexible \wrt array dimensions.
	
	\subsection{Unfolded PGD}
	In \cite{pellaco2020_deep_unfolding_pgd}, single-cell scenarios with single-antenna \glspl{ue} are exclusively studied. 
	Here, the application of \gls{kkt} conditions (as in \eqref{eq:theory:dualfun} and \eqref{eq:theory:vupdate}) is avoided by computing the subproblem for $\matr{V}$ using a finite number $Q$ of sub-iterations of \gls{pgd} on the \gls{wmmse} objective \eqref{eq:main:reform_sum_rate_problem} \wrt $\matr{V}$.
	The \gls{pgd} step-sizes $\gamma_{\ell q}$ are chosen as trainable parameters, where $q=1,\dots,Q$ is the \gls{pgd} sub-step index within an outer \gls{wmmse} layer $\ell$.
	This particular unfolding can be straightforwardly extended to multicell scenarios, as alluded to by the original authors.
	Furthermore, we extend the approach to the case where $N_i > 1$ as opposed to $N_i=1$ as considered by the original authors, leading to the subiteration update $\hat{\matr{V}}_i\itindex{\ell,q} = \matr{V}_i\itindex{\ell,q-1} - \gamma_{\ell q} \left(\matr{R}_k\itindex{\ell} \matr{V}_i\itindex{\ell,q-1} - \vmtilde_i\itindex{\ell}\right)$ where $\matr{V}_i\itindex{\ell,0} = \matr{V}_i\itindex{\ell-1}$.
	Each substep is succeeded by an Euclidian projection \eqref{eq:main:pownorm} to the feasible set.
	\begin{table*}
		\centering
		\caption{Relative \gls{wsr} of the Unfolded PGD in multicell MU-\gls{mimo} compared to GCN-WMMSE (proposed).} \label{tab:relatedwork:mimo_pgd}
		\begin{tabularx}{0.85\linewidth}{Xrrrrrr}
			\toprule
			Scenario & \multicolumn{4}{c}{Unfolded PGD $\rate_\Sigma^\mr{rel.}$ (\%)} & GCN-WMMSE $\rate_\Sigma^\mr{rel.}$ (\%) & WMMSE50 $\rate_\Sigma$ (\si{\nats})\\
			\midrule
			& $Q=4$ & $Q=8$ & $Q=12$ & $Q=16$ & & \\\cmidrule(lr){2-5}
			I: Simple Rayleigh & 69.71 & 77.87 & 80.77 & 81.84 & 89.21 & 25.22\\
			\midrule
			& $Q=16$ & $Q=24$ & $Q=32$ & $Q=40$ & & \\\cmidrule(lr){2-5}
			II: Scenario of \tab{tab:main:exp_base_scenarioparam} & 46.30 & 48.88 & 48.88 & 52.31 & 94.26 & 92.56\\
			\bottomrule	
		\end{tabularx}
		\vspace{-10pt}%
	\end{table*}
	\par
	We compare the achievable rate obtained by the unfolded \gls{pgd} \gls{wmmse} to the proposed GCN-WMMSE. %
	The GCN-WMMSE networks are trained using the same hyperparameters as in \tab{tab:main:exp_base_modelparam}, while the unfolded \gls{pgd} \gls{wmmse} is trained for $T=20000$ steps with a multiplicative step size decay of 0.1 after $5000$ steps each and $\mathcal{L}=\matset{1,\dots,L}$. 
	All networks are configured for $L=7$ layers. %
	In Scenario I, the networks are trained and validated on scenarios with $K=4$ \glspl{bs} with $M=4$, and $I=8$ \glspl{ue} uniformly assigned with $N=2$. 
	Channel matrix coefficients are drawn from $\cnormdistr{0,1}$ and we set $\sigma_\mr{UE}^2=\SI{0}{\dBm}$ and $P_\mr{BS}=\SI{20}{\dBm}$.
	\par
	\tab{tab:relatedwork:mimo_pgd} (Scenario I) demonstrates that the unfolded \gls{pgd} \gls{wmmse} model achieves $81.84\%$ relative \gls{wsr} with $16$ \gls{pgd} substeps, clearly saturating \wrt the number of sub-steps $Q$. 
	In comparison, the proposed GCN-WMMSE achieves $89.21\%$. 
	We remark, however, that this comes at the cost of an eigendecomposition of $4\times 4$ matrices in \eqref{eq:theory:compslack}, which significantly contributes to the total computational cost.
	\par
	On the other hand, when considering scenarios as in section \ref{sec:main:generalization} with parameters as in \tab{tab:main:exp_base_scenarioparam}, the unfolded \gls{pgd} fails to achieve a relative \gls{wsr} above $53\%$ for $Q=40$ subiterations.
	Higher required values for $Q$ are expected as $M$ increases, however, a saturation in performance is evident. 
	This can be attributed to reciprocal interference alignment \cite{gomadam2008_interference_alignment, schmidt2009_wmmse_miso} which leads to the eigenvalues of $\matr{R}_k\itindex{\ell}$ becoming more dissimilar with increasing iteration index $\ell$, \gls{snr} or degrees of freedom of the scenario. %
	A badly conditioned matrix can drastically decrease the convergence speed of a first-order \gls{qcqp} optimization methods such as \gls{pgd} \cite[Th. 11]{necoara2016_linearconvergence}.
	Therefore, globally trained \gls{pgd} step sizes are too large and do not accurately converge towards the local optimum.
	In comparison, \eqref{eq:theory:dualfun} and \eqref{eq:theory:vupdate} correspond to a second-order method.
	
	\subsection{IAIDNN}
	The authors of \cite{hu2021_iadnn} consider single-cell MU-\gls{mimo} scenarios.
	In this limited case, the power constraint is guaranteed to be exactly met.
	Hu \etal consequently construct an equivalent objective to the \gls{wsr} maximization by absorbing the constraint into the objective function, thereby developing a \gls{wmmse}-like algorithm without Lagrangian dual variables and eigendecompositions.
	Within the architecture based on unrolling this algorithm, that is termed iterative algorithm-induced deep neural network (IAIDNN), matrix inverse multiplications, \eg, for a matrix $\matr{B}_i\inv \vmtilde_i$, are approximated by the operation
	$
		\matr{B}_i\inv \vmtilde_i \approx \left(\matr{B}_i^+ \matr{X}_i^b + \matr{B}_i \matr{Y}_i^b + \matr{Z}_i^b \right)\vmtilde_i + \matr{O}_i^b %
	$
	which is derived from the first-order Taylor approximation of the inverse.
	The matrix $\matr{B}_i^+$ is the inverse diagonal of $\matr{B}_i$.
	Note that the conformable parameter matrices $\matr{X}_i^b$, $\matr{Y}_i^b$, $\matr{Z}_i^b$ and $\matr{O}_i^b$ are separate for every \gls{ue} index $i$.
	Furthermore, the $\matr{V}$-step in the last layer remains exempt from the approximation, which is empirically critical to achieve a high \gls{wsr}.
	The authors set $\matr{Y}_i^b = 0$ for all $i$ in practice. %
	For more details, see \cite{hu2021_iadnn}.
	Although somewhat similar to the proposed GCN-WMMSE architecture on first glance, the GCN-WMMSE network components are motivated by graph filters and feature augmentations such as nonlinearities and skip connections.
	Additionally, it is fully general with any given parameter set and is suitable for multicell scenarios, while a particular parameter set limits IAIDNN in its number of \glspl{ue} or number of supported antennas. 
	\begin{table}
		\centering
		\caption{Performance of IAIDNN compared to GCN-WMMSE (proposed) on single-cell scenarios with $P_\mr{BS}=\SI{20}{\dBm}$, $\sigma^2=\SI{0}{\dBm}$ and $N=2$. * indicates the results reported in \cite{hu2021_iadnn}.} \label{tab:relatedwork:iadnn_perf}
		\begin{tabularx}{\linewidth}{Xlrrr}
			\toprule
			\multicolumn{2}{l}{Scenario $(M, I)$} & $(8, 4)$ & $(16, 8)$ & $(32, 16)$\\
			\midrule
			Abs. Rate $\rate_\Sigma$ (\si{\nats})& WMMSE50 & 32.22 & 58.53 & 112.08 \\
			\midrule
			\multirow{5}{*}{Rel. Rate $\rate_\Sigma^\mr{rel.}$ (\%)} & IAIDNN & 85.35 & 90.52 & 91.93\\
			& IAIDNN Imp. & 82.31 & 86.49 & 91.38\\
			& IAIDNN SP & --- & 87.29& ---\\
			& *IAIDNN \cite{hu2021_iadnn} & *91.35 & *92.13 & *92.63\\
			& GCN-WMMSE & 96.34 & 101.66 & 101.65\\
			\bottomrule	
		\end{tabularx}
	\end{table}
	\begin{table*}
		\centering
		\caption{Average \gls{ue} rates for a \gls{mimo} \acrlong{bch} with $M=16$ and $I=8$ in \si{\nats} for WMMSE MRC, IAIDDN and IAIDDN with parameters shared across \glspl{ue} (IAIDNN SP).}
		\label{tab:relwork:iadnn_userrates}
		\begin{tabularx}{12cm}{Xrrrrrrrrr}
			\toprule
			\textbf{User} $i$ & \textbf{Sum} & 1 & 2 & 3 & 4 & 5 & 6 & 7 & 8\\
			\midrule
			\textbf{WMMSE MRC} & 58.25 & 7.40 & 7.38 & 7.15 & 7.22 & 7.23 & 7.29 & 7.30	& 7.28\\
			\textbf{IAIDNN} & 52.98 & 8.14 & 8.13 & 8.12 & 4.06 & 4.0 & 4.15 & 8.14 & 8.16\\
			\textbf{IAIDNN SP} & 51.10 & 6.39 & 6.41 & 6.34 & 6.38 & 6.47 & 6.37 & 6.33 & 6.41\\
			\bottomrule
		\end{tabularx}
		\vspace{-10pt}%
	\end{table*}
	\par
	We reimplemented the IAIDNN architecture in PyTorch, taking advantage of automatic differentiation and ADAMW optimizers instead of a manual gradient computation as in \cite{hu2021_iadnn}.
	The IAIDNN networks are trained with a step size of $10^{-3}$, which is decayed after $2500$ steps each, beginning at $t=5000$.
	The IAIDNN hyperparameters are tailored to the scenario configurations while the GCN-WMMSE networks adopt the parameters from \tab{tab:main:exp_base_modelparam}.
	The number of network layers is $L=7$.
	Only single-cell scenarios are considered and channel matrix coefficients are sampled from $\cnormdistr{0,2}$ as in \cite{hu2021_iadnn}.
	\par
	As \tab{tab:relatedwork:iadnn_perf} demonstrates, the proposed GCN-WMMSE networks outperform the IAIDNN networks by $9\%$ (compared to results reported in \cite{hu2021_iadnn}) for $M=16$ and $32$ with the same number of layers, however, note the computational cost of the eigendecomposition. %
	We were unable to reproduce the results reported in \cite{hu2021_iadnn} exactly, especially regarding the improved IAIDNN architecture which applies matrix inversions.
	We further remark that an IAIDNN network produces degenenerate asymmetric results on symmetric data sets.
	Specifically, some \gls{ue} indices are on average disadvantaged \wrt their \gls{ue} rate $\rate_i$, as \tab{tab:relwork:iadnn_userrates} demonstrates.
	This can be attributed to using different network parameter groups per \gls{ue} index $i$, leading to some receive signal spaces of some receivers being permanently assigned to interference.
	Employing shared parameters (SP) instead leads to balanced rates between indices $i$ while also reducing the number of trainable parameters and reducing the computational complexity.
	However, the \gls{wsr} slightly reduces due to slower convergence since interference allocation is not a priori anymore.
	This proves the importance of ensuring equivariance for machine learning models in such optimization problems.
	
	\section{Conclusion}\label{sec:conclusion}
	We propose a distributed unrolled architecture based on the classical WMMSE algorithm, termed GCN-WMMSE, which is applicable to multicell MU-MIMO wireless networks.
	Although the computation complexity per iteration is of the same order as the original \gls{wmmse} algorithm, the number of iterations is massively reduced while achieving a similar rate as the WMMSE, decreasing the communication overhead in a distributed deployment.
	For the same set of parameters, it maintains its performance across changing scenario configurations in most instances, proving its excellent generalization capabilities.
	Additionally, it compares favorably to previous unrolled WMMSE architectures despite its generality. 
	Future investigations could address suitable approximate solutions of the downlink beamformer subproblem in multicell \gls{mu}-\gls{mimo} networks.
	Furthermore, it must be noted that the instantaneous capacity model has practical limitations and the more appropriate ergodic channel model is subject to subsequent research.
	\printbibliography
\end{document}